\documentclass[sn-mathphys,Numbered]{sn-jnl}

\usepackage[caption=false]{subfig}
\usepackage{graphicx}%
\usepackage{multirow}%
\usepackage{amsmath,amssymb,amsfonts}%
\usepackage{amsthm}%
\usepackage[capitalise]{cleveref}
\usepackage{mathrsfs}%
\usepackage[title]{appendix}%
\usepackage{xcolor}%
\usepackage{textcomp}%
\usepackage{manyfoot}%
\usepackage{booktabs}%
\usepackage{algorithm}%
\usepackage{algorithmicx}%
\usepackage{algpseudocode}%
\usepackage{listings}%

\theoremstyle{definition}%

\newtheorem{example}{Example}
\newtheorem{lem}{\textbf{Lemma}}
\newtheorem{corr}{\textbf{Corollary}}
\newtheorem*{prrf}{\textbf{Proof}}
\newtheorem*{observation}{\textbf{Observation}}

\begin{document}

\title[Article Title]{Mitigating Procrastination in Spatial Crowdsourcing Via
Efficient Scheduling Algorithm}

\author*[1]{\fnm{Naren} \sur{Debnath}}\email{nd.20cs1104@phd.nitdgp.ac.in}

\author[2]{\fnm{Sajal} \sur{Mukhopadhyay}}\email{smukhopadhyay.cse@nitdgp.ac.in}
\equalcont{These authors contributed equally to this work.}

\author[3]{\fnm{Fatos} \sur{Xhafa}}\email{fatos@cs.upc.edu}
\equalcont{These authors contributed equally to this work.}

\affil[1,2]{\orgdiv{Department of Computer Science and Engineering}, \orgname{National Institute of Technology
Durgapur}, \orgaddress{\city{Durgapur}, \postcode{713209}, \state{West Bengal}, \country{India}}}

\affil[3]{\orgdiv{Department of Computer Science}, \orgname{Universitat Politècnica de
Catalunya}, \orgaddress{\city{Barcelona}, \postcode{08034}, \state{Catalonia}, \country{Spain}}}

\abstract{Several works related to spatial crowdsourcing have been proposed in the direction where the task executers are to perform the tasks within the stipulated
deadlines. Though the deadlines are set, it may be a practical scenario that
majority of the task executers submit the tasks as late as possible. This
situation where the task executers may delay their task submission is termed
as procrastination in behavioural economics. In many applications, these
late submission of tasks may be problematic for task providers. So here, the
participating agents (both task providers and task executers) are articulated
with the procrastination issue. In literature, how to prevent this procrastination within the deadline is not addressed in spatial crowdsourcing scenario. However,
in a bipartite graph setting one procrastination aware scheduling is proposed
but balanced job (task and job will synonymously be used) distribution in
different slots (also termed as schedules) is not considered there. In this
paper, a procrastination aware scheduling of jobs is proliferated by proposing an (randomized) algorithm in spatial crowdsourcing scenario. Our algorithm ensures that balancing of
jobs in different schedules are maintained. Our scheme is compared with the
existing algorithm through extensive simulation and in terms of balancing
effect, our proposed algorithm outperforms the existing one. Analytically it
is shown that our proposed algorithm maintains the balanced distribution.}


\maketitle

\section{Introduction}\label{intro}
Consider a logo designing task of a multinational company which can be distributed to a set of current employees of the company or to obtain the best design, the company may float the task worldwide. Anyone possessing the design capacity can contribute her design to that company no matter which corner of the world she belongs to. This situation for asking contribution from unbounded bulk people  around the world is generally termed as \emph{crowdsourcing}  as the task has been outsourced from the crowd without categorisation \cite{howe2006rise}\cite{alam2020crowdsourcing}\cite{cricelli2021crowdsourcing}\cite{castillo2022crowdsourcing}\cite{pan2011survey}. This concept has been utilized since 2006 and proliferated in various directions from disaster management \cite{besaleva2013crowdhelp} to atmospheric sciences \cite{muller2015crowdsourcing}, from agriculture to logistics \cite{han2020exact}, IoT to smart cities \cite{singh2020budget} and there are many to name.

Consider another scenario where an agency wants to measure the air moisture level in multiple agricultural areas in a city or in a district or so. In present time there are various sensory devices that can measure the  moisture level. Also, with the increased usage of mobile devices (smartphones) \cite{wang2020promoting}, mobile users can easily read those sensory data using appropriate mobile application or through sensors in mobile phones \cite{yang2012crowdsourcing}. To perform this kind of task from multiple locations, the agency could distribute the task to the general crowd who have smartphones. The extension of crowdsourcing, specifically crowdsensing through mobile devices (collecting sensor data) for a particular location, is termed as mobile crowdsourcing \cite{fuchs2014architecture}\cite{wang2016quacentive}. Mobile crowdsourcing (MCS) is emerging since 2009 with mobile applications like \emph{m-Clerke} \cite{yan2009mcrowd}, a mobile (iPhone) application which utilizes sensors in iPhone to perform various crowdsourcing tasks, \emph{txteagle} \cite{eagle2009txteagle} through which agents (henceforth, task executers and task provider or requesters will also be termed as agents) can perform the task of \emph{translation, transcription and surveys}. There are many other researches which has given MCS immense popularity and utmost importance \cite{phuttharak2018review}.

When MCS tasks are to be executed at different geographic areas with spatiotemporal constraint, it is termed spatial crowdsourcing (SC),
which is the boiling research topic and also the subset of MCS \cite{tong2020spatial}\cite{gummidi2019survey}\cite{wang2021worker}\cite{hamrouni2020spatial}\cite{zhang2016spatialrecruiter}. Some of the most customary examples of SC application are food delivery service\cite{liu2018foodnet}, on-demand taxi\cite{guo2018task} and bike service \cite{gummidi2019survey}. SC is also applied for collecting geospatial data \cite{wang2021worker}, traffic information \cite{wang2017secure}, pollution\cite{zhang2016reliable}, disaster \cite{tehrani2018toward}\cite{kong2019mobile}, journalism\cite{li2021toward} and many more to name with requirement of spatial information. Though the demand and application of SC are increasing incrementally, some loopholes have been found, such as acquisition of inaccurate data, bounded capability of sensing devices, or untruthful data reporting from the worker end, etc., which has been notified and addressed in \cite{zhang2019expertise}.

At an abstract level in SC\footnote{applicable in crowdsourcing in general sense}, we have task provider(s) and task executers where the task provider(s) disseminate the tasks to the task executers. Many researches related to SC have been carried out in the direction where the task provider is constrained by deadlines \cite{yadav2021budget}\cite{tong2020spatial}\cite{djigal2022buda}\cite{fu2021fairness}\cite{deng2013maximizing}\cite{mukhopadhyay2021egalitarian}\cite{singh2020budget}\cite{ma2021mixed}. Even if the deadlines are set, it may be a realistic situation that most of the agents submit the tasks at the penultimate day or at the last day of submission. This situation where the agents may delay their task submission is termed as \emph{procrastination} \cite{roughgarden2016cs269i}\cite{kleinberg2014time}\cite{tang2017computational}\cite{kleinberg2017planning}\cite{gravin2016procrastination} in behavioural economics\cite{akerlof1991procrastination}\cite{kahneman2013prospect}\cite{kleinberg2014time}\cite{tang2017computational}. In many applications, these late submissions of tasks may be problematic for task providers. 
In the SC literature (also in crowdsourcing), how to prevent this procrastination within the deadline is not addressed. However, in a bipartite graph setting, one procrastination aware scheduling is proposed in \cite{wang2019procrastination}. In \cite{wang2019procrastination}, balanced job distribution in different slots (also termed as schedules) regarding procrastination perspective is not considered. In this paper, an algorithm named, \emph{Procrastination Preventive Scheduling of Jobs: A Balanced Perspective (PPSJBP)} is proposed in SC scenario (also applicable in crowdsourcing), so that balancing of jobs in different schedules are maintained (\emph{The detailed objective of the paper is presented in 
\cref{obj}}). We have compared our scheme with \cite{wang2019procrastination} and have observed that in terms of balancing effect, our proposed algorithm outperforms \cite{wang2019procrastination}. Also we have shown analytically that our proposed algorithm maintains the balanced distribution.

\section{Literature Review}
 \label{csbe}
In \cite{roughgarden2016cs269i}\cite{kleinberg2014time}\cite{akerlof1991procrastination}, an interesting story of the famous economist George Akerlof is presented, which goes like this.
During his visit to India for one year,
he had to return some of his friend’s luggage to USA. Every day Akerlof thought he would send off
the luggage, but he delayed one more day due to the lengthy
transportation process in India (he delayed as he thought that the loss of a whole day might be costly for him). This thought process of
sending off the boxes on next day was continued for a few
months before he finally managed to send the luggage. This story shows that procrastination could be a natural process, and we can build an efficient system taking procrastination into consideration. This viewpoint has been the prime motivation to write our paper.

Another concept of behavioral economics is loss aversion, where the participating agents perceive loss more seriously than the equivalent profit. Some works based on this aspect (loss aversion) of behavioral economics have been done in the Mobile Crowdsourcing and Spatial Crowdsourcing scenarios.

In spatial crowdsourcing, according to the agents' participating location, there are two types of regions that are marked as \emph{higher participation} and \emph{sparse participation} regions. The task completion rate in \emph{sparse} regions is much slower than in \emph{higher participation} regions. \emph{Liu et al.} in their paper \cite{liu2020incentive} have addressed this issue by proposing two incentive mechanisms based on behavioral economics that would leverage to the selection and payment for participants (agents) for sparse regions. These two mechanisms are - \emph{Incentive mechanism based on Behavioral Economics for Participant Selection (IBE-PS)} and \emph{Incentive mechanism based on Behavioral Economics for Payment Decisions (IBE-PD)}. The basis of the first mechanism is the reference effect and the second one is the loss aversion. They have shown that these two mechanisms have effectively enhanced the participation of agents at \emph{sparse} regions by which the loss of utility is diminished and the task completion rate is enhanced.

In \cite{li2019crowdsensing}, we found another stress on behavioral economics where two problems are addressed. One is the \emph{fungibility} of utility of various tasks and the other is the \emph{consistent} behavioural preferences of the users (agents). According to the traditional economic theory, the utility of different tasks for agents is indistinguishable which refers to the term \emph{fungible}. Mental Accounting in \cite{thaler1999mental} first unveiled the fact that the utility of different tasks are different, $i.e.,$ non-fungible and users' behavioral preferences are not consistent. Based on Mental Accounting, \emph{Deng et. al} in \cite{li2019crowdsensing} proposed an incentive mechanism called \emph{Mental Accounting Auction Incentive Mechanism (MAAIM)}. The proposed incentive mechanism is characterised by reference dependence, loss aversion, and sensitivity decline. Reference dependence and sensitivity decline have created an \emph{external reference environment} and \emph{internal reference point} which motivated the agents participating in the system. As a result, the utilisation of sensing platform and involvement of agents in task participation are improved. They have designed a payment mechanism based on loss aversion which also motivates the agents to collect better quality of data.

\emph{Li et al.} in \cite{li2018analysis} have proposed a scalable payment algorithm for agents based on the parameter called loss aversion.  In their work, the loss aversion is introduced as a coefficient (parameter) in the utility function (for calculation of agents' utility) in order to modify the payment of agents to motivate their cooperative behaviour which leads to the enhancement of the crowdsensing platform.

Apart from crowdsourcing, there are other areas where the mechanisms are proposed considering the behavioral aspects of the agents who are participating in the system. One such work is presented in \cite{liu2021cooperation}. In their work, an incentive mechanism is proposed based on loss aversion called \emph{Loss Aversion Incentive Mechanism (LAIM)} for \emph{Vehicular Ad-Hoc Networks (VANET)}. The incentive mechanisms for \emph{VANET} until \cite{liu2021cooperation} are based on expected utility theory of traditional economics which assumes that participating in a system, the agents perceive the same effect by observing the same amount of gain or loss. However, the most realistic scenario would be to design mechanisms in \emph{VANET} that will take the behavioral aspect of the agents where the agents perceive the profit and loss differently. 
In \cite{liu2021cooperation}, a mechanism is proposed by assuming that the nodes will have this behavioral aspect in consideration and with that they have shown, their mechanism will improve the cooperation among the participating nodes.

As stated earlier, when the agents are executing the time bound jobs, it is quite realistic that they may submit the tasks as late as possible and these late submission of tasks may be problematic for task
requester. The problem of the task provider aggravates if the task provider is also constrained by deadline. The above review 

has not portrayed one very important step (often ignored), which is, the task provider is also constrained by deadline. The problem that arises, if the task provider is constrained by deadline, could be understood with the following example. 

Consider a faculty member announces 3 assignments to the students be submitted within $27^{th}$ of a month and the marks of the submitted assignments which is to be given within $30^{th}$ of the same month by the faculty member. In worst case, all the assignments are submitted to her on $27^{th}$. Thus, the evaluation quality will be compromised as the time left for evaluation is only $3$ days which is much less for evaluation in worst case scenario. So, the evaluation quality in this case is diminishing as the faculty member is getting less time to finish the stipulated work. This situation arises due to the fact that the deadline for submission of the assignments was not designed efficiently \emph{i.e.,} a single big deadline is given to the students to submit all the assignments. To mitigate this issue related to the deadline, a work is presented in \cite{wang2019procrastination} where they have proposed a Procrastination-aware-Scheduling algorithm (the algorithm is termed as \emph{OffPSP}).

This algorithm has allocated all jobs into the schedules in some sorted order based on a threshold, set for each schedule. Accordingly, this algorithm has produced a set of unbalanced schedules in terms of cost and number of jobs allocated.  With these unbalanced schedules, the agent may face the same problem (will not get ample time) as stated earlier. In our paper, the issue related to this unbalancing effect on the schedules is addressed.

\section{System Models}\label{sysmod}
\subsection{Preliminaries:}\label{prel}
Consider a 3-week short term course and the organiser of the course wants that a participant in the course, needs to submit two assignments at the end of the short term course. It may not be a good schedule, if we consider the fact that the bulk of the participants will procrastinate and submit the assignments as late as possible within the deadline (in our example, it is three weeks).

In this case, to prevent the procrastination, the organiser can reschedule the submission process to ensure a smooth and consistent submission from the participant. Here, the 3-week duration can be decomposed into 3 unit intervals - first week, second week, and third week and they can assure that one assignment to be completed by the end of first week or second week and the next assignment could be solved in the third week. In literature, this is called the choice reduction \cite{kleinberg2014time}\cite{roughgarden2016cs269i}. This viewpoint of choice reduction is considered while designing our proposed algorithm.

\subsection{Notations and their Usage}\label{moddef}
\textbf{Schedules:} In this model, a time horizon $T$ is divided into a set of schedules $\Delta=\{\delta_1,\delta_2,\cdots,\delta_l\}$ and the time duration of each $\delta_i$ is $\tau_i$. Here $\tau_1=\tau_2=\cdots=\tau_{|\Delta|}$ and $|\Delta|$ is defined as $|\Delta|=\frac{T}{\tau_i},\ where\ \tau_i\in \mathbb{R}, \tau_i\ne0,\ and\ \tau_i\le T$. It is to be noted that, $\tau_i$ depends on the applications we are working in and thus our algorithm is flexible enough to be amenable to any such application. As $T$ is divided into several schedules and each schedule has a time duration $\tau_i$, we can write $T=\{\tau_1,\tau_2,\cdots,\tau_l\}$. So, each schedule $\delta_i$ is a tuple denoted as $\delta_i=\{\delta_i^{w}, \tau_i\}$, where $\tau_i\in T$.

\noindent\textbf{Location and Task:} An entire SC zone is divided into a set of locations $\mathcal{L}=\{\mathfrak{l}_1,\mathfrak{l}_2,\cdots,\mathfrak{l}_m\},\ where\ m=|\mathcal{L}|$. Each location $\mathfrak{l}_i$ is populated with a set of jobs (dispensed by an agent) which could be denoted as $\Gamma=\{\gamma_1,\gamma_2,\cdots,\gamma_n\}$ where $n=|\Gamma|$. Each job $\gamma_i$ is also associated with a cost $\chi_i$. The cost of jobs may be considered as time span, amount of internet expenses, battery power, or distances to be travelled, etc. as per the SC system. As $\gamma_i$ is associated with a cost $\chi_i$ and a location $\mathfrak{l}_i$, we can redefine $\Gamma$ more formally as follows: $\Gamma=\{<\gamma_1,\chi_1, \mathfrak{l}_i>,<\gamma_2,\chi_2, \mathfrak{l}_i>,\cdots,<\gamma_n,\chi_n, \mathfrak{l}_i>\}$, where the triplet $<\gamma_i,\chi_i,\mathfrak{l}_i>$ denotes the job identifier, the cost associated with it and the location at which the job is to be executed respectively.
These jobs are to be scheduled into each of $\delta_i$ ($\delta_i\in \Delta$). Each $\delta_i$ is comprising of a set of jobs denoted as 
$\delta_i^{w}:\{\gamma_j\mid \gamma_j\in \Gamma\setminus \bigcup_{\substack{i^{'} =1 \\ i^{'} \neq i}}^l \delta_{i^{'}}^{w}\},\ \delta_i^{w} \subseteq \Gamma$. 
As a ready reference, all symbols used so far in this paper are listed with their brief description in \cref{ujcs-tab0}.
\begin{table}
 \caption{Important symbols and their definitions}\label{ujcs-tab0}
 \centering
     \begin{tabular}{ll}
     \hline                              
         Symbol        & Definition                        \\
       \hline
       $\Gamma$  &   Set of jobs \\
       $\mathcal{L}$ & Set of locations\\
       $\Delta$  &   Set of schedules \\
       $\gamma_i$    &   $i^{th}$ job in set $\Gamma$\\
       $\mathfrak{l}_i$ & $i^{th}$ location in set $\mathcal{L}$\\
       $\delta_i$    &   $i^{th}$ schedule in set $\Delta$\\
       $n$   &   Number of jobs\\
       $l$   &   Number of schedules\\
       $T$   & Total time horizon      \\
       $\tau_i$  &   Time duration of each schedule\\
       $\delta_i^w$  &  Set of jobs allocated to $\delta_i$\\
       $K$   &   Number of iterations    \\
       $k$   &   An arbitrary iteration  \\
       \hline                                                   
     \end{tabular} 
    
\end{table}
\subsection{Objective}\label{obj} 
Given a set of jobs $\Gamma$ and the set of schedules $\Delta$,  jobs are to be allocated into several schedules $\delta_i$ ($\delta_i\in \Delta$) in a balanced way for an agent (or for the set of agents who will perform the same set of jobs) such that the following constraint is satisfied. \[\min_{k}\{Var(\Delta_k)\}\] where $k$ is the iterative index of $K$ iterations and $Var(\Delta_k)$ is represented as the variance of the set of schedules ($\Delta$) at $k^{th}$ iteration. By balancing we mean, the variance in $\Delta_k$ will be least. 

The objective of our mechanism is to prevent procrastination through allocation of jobs into a set of schedules in balanced manner. When all schedules are balanced, all jobs with higher cost will uniformly be distributed into different schedules. In such a way, the accumulation of higher cost jobs into one schedule will be prevented. Thus, delaying in execution of these higher cost jobs along with the other jobs allocated into the same schedule could be prevented. Another aspect we have considered here that as we have divided the large time horizon $T$ in several schedules (refer \cref{moddef}), the choices of procrastination are getting reduced. 
In the similar line we have stated earlier, we can argue that if all jobs are given $2$ weeks time $i.e.$, all the jobs can be completed any time within those $2$ weeks, chances of procrastination may be more. Instead,    if the $2$ weeks time horizon is divided into $2$ different schedule with $1$ week time then the choices for submission of jobs are reduced. Here we have generalized the model by dividing the entire time horizon in $l$ number of divisions and thereby reducing the choices substantially.

We have developed our algorithm by considering an arbitrary ($i^{th}$) location so that it could be applicable to all $m$ locations under consideration in this SC model.

\begin{figure}
\centering
    \includegraphics[width=\linewidth]{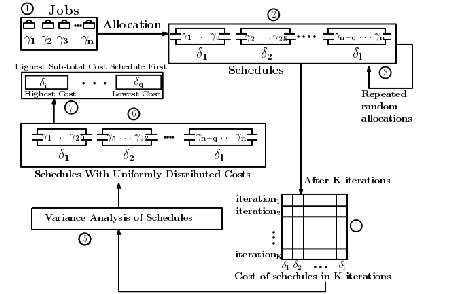}
    \caption{Flow Diagram of the proposed mechanism}
    \label{flow}
\end{figure}
\section{Proposed Mechanism}\label{ucjs}
This section discusses the proposed algorithm (PPSJBP) in two different ways. First, a schematic diagram of the proposed mechanism is provided. Second, the step-wise algorithm is discussed. Finally,  an example with an arbitrary set of data is furnished at the end of this section to illustrate the workflow of our algorithm.

\subsection{Schematic Diagram of PPSJBP}\label{scdp} \cref{flow} shows the flow of our proposed mechanism. A pool of jobs is available to the platform (block 1). Our mechanism randomly assigns those jobs into $|\Delta|$ schedules (block 2). To achieve higher accuracy in balanced distribution of jobs into schedules, the random distribution of $|\Gamma|$ jobs is repeated for $K$ times (block 3). For each $k\in K$, the variance of $\Delta_k$ is calculated (block 4). Let us denote the variance of $\Delta_k$ by $Var(\Delta_k)$. For balancing purpose, we need to calculate $\min_{k}\{Var(\Delta_k)\}$ (block 5). We obtain an almost balanced set of schedules as an outcome of block 5 (shown in block 6). Finally selected schedules are presented to the agents in non increasing order (block 7) of their cost. It is to be noted that our algorithm will also work for heterogeneous set of agents. By heterogeneous we mean that the set of jobs to be executed by them are different. In this case, we can apply our algorithm separately to each agent.

\begin{figure}
\centering
    \includegraphics[width=\linewidth]{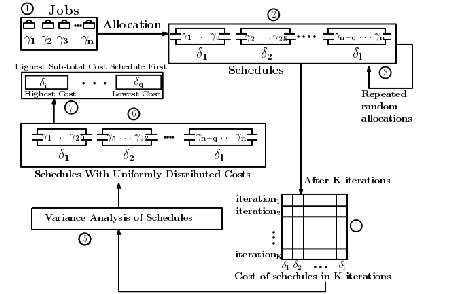}
    \caption{Flow Diagram of the proposed mechanism}
    \label{flow}
\end{figure}

\subsection{Procrastination Preventive Schedule of Jobs: A Balanced Perspective (PPSJBP)}\label{algo}
The glimpse of the mechanism has already been discussed in \cref{sysmod}. The detail method in the form of an algorithm is explained in this subsection. Our algorithm has four main sections.

\begin{itemize}
    \item \emph{Main Routine}
    \item \emph{Repeated Random Allocation (RRA)}
    \item \emph{Most Balanced Distribution Find (MBDF)}
    \item \emph{Largest Cost Schedule First (LCSF)}
\end{itemize}

\begin{algorithm}
\caption{Main Routine}\label{alg:one}
\begin{algorithmic}[1]
    \Require Set of jobs with their costs: $\Gamma=\{<\gamma_1,\chi_1>,<\gamma_2,\chi_2>,\cdots,<\gamma_n,\chi_n>\}$
    \Ensure Most Balanced set of schedules of jobs: $\Delta=\{\delta_{1},\delta_{2},\cdots,\delta_{l}\}$
    \State $S_{Tot},\ K \gets RRA(\Gamma)$
    \State $index \gets MBDF(S_{Tot},K)$
    \State $MOD\_SCHDL \gets LCSF(index,1,s\_l)$
    \State $End$
\end{algorithmic}
\end{algorithm}

\subsubsection{Main Routine}
The main routine calls three subroutines, the first one is 
\emph{Repeated Random Allocation (RRA)} which is framed in \cref{alg:two}, allocates jobs to the set of schedules and calculates the total cost of jobs in each schedule through $K$ repeated allocation strategy. The input to this subroutine is the set of jobs $|\Gamma|$. The output of this subroutine is a cost matrix where every row represents the $i^{th}$ iteration and each column represents the total cost of a schedule at that iteration.

The second one is \emph{Most Balanced Distribution Find (MBDF)} subroutine, given in \cref{alg:three} which finds the most balanced set of schedules (in terms of total cost) using variance analysis. The input to this subroutine is the cost matrix returned by \emph{RRA} subroutine.

The third subroutine is \emph{Largest Cost Schedule First (LCSF)} which reorders the schedules according to their total costs and the corresponding algorithm is presented in \cref{alg:four}. The set of schedules (most balanced) returned by \emph{MBDF} is used as the input to this subroutine. All three subroutines are explained in detail from the next subsection.

\subsubsection{Repeated Random Allocation Algorithm}
In \cref{alg:two}, all necessary initialization have been done from line 1 to line 4. The total time frame is initialized in line 1 by subtracting the start time by finish time. The number of schedules is determined by dividing the total time frame by time unit ($t_{unit}$) in line 2. The time unit is same for all schedules and the value of which depends on the applications. A 2-D data structure is initialized in line 4 in which the total cost of each schedule obtained in each iteration will be stored.
In line 6, the outer loop starts which repeats the random allocation of jobs into different schedules for $K$ times, where $K$ is defined as a large number (we have considered 8000 and more for our simulation) in line 5. In line 7, a temporary job set is initialized which will repeatedly be used in all $K$ iterations. The inner loop starts in line 8 and continues until all jobs are allocated into schedules for a single iteration of the outer loop. Every job from the job set and the schedule into which a particular job will be allocated are selected randomly in line 9 and line 10 respectively. The randomly selected job is then assigned to the randomly selected schedule in line 11. The temporary job set is updated by removing the allocated job from that set at line 12. After a job is allocated into a schedule (at line 11), the  cost of the corresponding job is cumulatively added with the cost of other jobs in that schedule in line 13. Line 16 returns the total cost (of each schedule in each iteration) matrix to the main routine (line number 1 in \cref{alg:one}).

\renewcommand{\algorithmicrequire}{\textbf{procedure}}
\begin{algorithm}
\caption{Repeated Random Allocation}\label{alg:two}
\begin{algorithmic}[1]
\Require RRA($\Gamma$)
    \State $span \gets (F_t-S_t)$ \hspace*{\fill}      $\triangleright S_t:\ start\ time,\ F_t:\ finish\ time$
    \State $l \gets span/t_{unit}$      \hspace*{\fill}$\triangleright Number\ of\ schedules$
    \State $i\gets 0$
    \State $\hat{\Delta}\gets \{\}$   
    \State $K=\mathfrak{C}$ \hspace*{\fill}      $\triangleright \mathfrak{C}\ is\ a\ large\ number$
    \While{$i \leq K$}
       \State $\Gamma_{local} \gets \Gamma$
        \While{$|\Gamma_{local}| \geq 1$}
            \State $r \gets random(1,l)$
            \State $\hat{\Gamma} \gets random(\Gamma)$
            \State $\hat{\Delta}_r^i \gets \hat{\Delta}_r^i \cup \{\hat{\Gamma}\}$
            \State $\Gamma_{local} \gets \Gamma_{local}-\{\hat{\Gamma}\}$
            \State $sub\_tot_r^i \gets sub\_tot_r^i+cost(\hat{\Gamma})$
        \EndWhile
        \State $i\gets i+1$
    \EndWhile
    \State \Return $sub\_tot,\ K$
\end{algorithmic}
\end{algorithm}

\subsubsection{Most Balanced Distribution Find Algorithm}\label{mbdf}
In this algorithm, the set of schedules with minimum variance is obtained (most balanced set of schedules) by calculating and comparing the variances of the total cost matrix obtained in \cref{alg:two}. 
In the beginning of the algorithm, a variable \emph{min\_var} is initialized with $\infty$ which represents a large value, in line number 1. Line number 2 starts the outer level iteration for calculation of the variances and the minimum of them. From line number 3 to line number 7, the variance of each row of total cost matrix is calculated using the formula given in (\cref{eqn:1}), where, $L$ (in the algorithm, it is $|\Delta|$) is the number of samples and $\bar{\chi^*}$ ($i.e.,sub\_tot_{i}^{mean}$) is the sample mean of the sample set $\chi^*=\{\chi^*_1,\chi^*_2,\cdots,\chi^*_L \}$ (set of costs of all schedules in a particular iteration). In the algorithm, $\chi^*_i$ is the $sub\_tot_i^j$.
Line number 5 calculates the sum of squared difference between the cost of each schedule and the average cost in each iteration. Finally in line number 7, that sum is divided by $(L-1)$ to obtain the variance.
\begin{equation}\label{eqn:1}
    \nu_i =\frac{\Sigma_{i=0}^L (\chi^*_i-\bar \chi^*)^2}{L-1}
\end{equation}
From line number 8 to line number 11, the minimum of all variances is determined using the comparison and replace method, $i.e.$, the temporary variable \emph{min\_var} is compared to each variance and the variance lesser than the content of \emph{min\_var} replaces the current content of \emph{min\_var}. In line number 13, the index of the minimum variance, over all iterations, is returned to the main routine. This index calculation is mathematically represented as: $index=\operatorname*{arg\,min}_{ \forall \nu_i}\{\nu_i\}$.

\begin{algorithm}
\caption{Most Balanced Distribution Find}\label{alg:three}
\begin{algorithmic}[1]
\Require MBDF($sub\_tot$,$K$)
    \State $min\_var \gets \infty$
    \While{$i \leq K$}
        \State $sum \gets 0$
        \For{$j = 1$ to $|\Delta|$}
            \State $sum \gets sum+(sub\_tot_i^j-sub\_tot_{i}^{mean})^2$
        \EndFor
        \State $\nu_i\gets sum/(|\Delta|-1)$
        \If{$\nu_i \leq min\_var$}
            \State $min\_var \gets \nu_i$
            \State $index \gets i$
        \EndIf
    \EndWhile
    \State \Return $index$    
\end{algorithmic}
\end{algorithm}

\subsubsection{Largest Cost Schedule First Algorithm}
From MBDF algorithm (\cref{alg:three}) we have obtained the most balanced set of schedules with minimum variance. We go further from here by taking that set of schedules to our last algorithm where schedules will be published according to the order of their costs. To achieve that, the schedule with the highest total cost is published first. Then the schedule with second highest total cost is published and so on. This method is aligned in \cref{alg:four} (Merge Sort method is used here).
\begin{algorithm}
\caption{Largest Cost Schedule First}\label{alg:four}
\begin{algorithmic}[1]
\Require LCSF($index,1,|\Delta|$) 
        \If{$|\Delta|>1$}
            \State $mid \gets \frac{|\Delta|}{2}$
            \State $LCSF(sub\_tot^{index},1,mid)$
            \State $LCSF(sub\_tot^{index},mid+1,|\Delta|)$
            \State $merge(sub\_tot^{index},1,mid,|\Delta|)$   
        \EndIf
        \State \Return $sub\_tot^{index}$
\end{algorithmic}
\end{algorithm}

\textbf{Purpose of variance analysis:} The variance is a statistical method to observe the distance among the objects of similar type. In our proposed method, we are aiming to distribute jobs with various range of costs into different schedules so that all schedules get the jobs with balanced total cost. The variance analysis method is finding the set of schedules with minimum variance by calculating the variance of each set of schedules obtained in each iteration and then by comparing all the variances of all such sets. 
In \cref{ujcs-ex2}, the entire algorithm is presented with more exposure to the core aspect of our method.

\begin{table}
\caption{Sample Table for Repeated Random Job Allocation}
\label{ujcs-tab1}
\centering
 \begin{tabular}{ccccccc}
\hline
 \multirow{2}{*}{Iteration\#}& \multicolumn{3}{@{}c@{}}{Job Allocation($\mathbf{\gamma_i}$) in Schedules} & \multicolumn{3}{@{}c@{}}{$\mathbf{\Sigma \chi}$} 
 \\\cmidrule(l{5pt}r{5pt}){2-4}\cmidrule{5-7}
&$\delta_1$&$\delta_2$&$\delta3$&$\delta_1$&$\delta_2$&$\delta3$\\
\hline
1 & $\gamma_1$,$\gamma_3$,$\gamma_6$ & $\gamma_4$,$\gamma_5$&$\gamma_2$&27&10&2 \\ 
 
2 & $\gamma_1,\gamma_3$&$\gamma_4,\gamma_5$&$\gamma_2,\gamma_6$&12&10&17\\

3 & $\gamma_1,\gamma_5$&$\gamma_6$&$\gamma_2,\gamma_3,\gamma_4$&13&15&11\\
 
4 & $\gamma_1,\gamma_3$&$\gamma_4,\gamma_5,\gamma_2$&$\gamma_6$&12&12&15\\
 
5 & $\gamma_1,\gamma_3,\gamma_2$&$\gamma_4,\gamma_5$&$\gamma_6$&14&10&15\\
 \hline
 \end{tabular}

\end{table}
\begin{example}\label{ujcs-ex2}
An example in this subsection is presented to comprehend the algorithm using some arbitrary data. Let there be 6 jobs in $\Gamma\gets\{<\gamma_1,4>, <\gamma_2,2>, <\gamma_3,8>, <\gamma_4,1>, <\gamma_5,9>, <\gamma_6,15>\}$. Let $t_{unit}\gets 300,\ S_t\gets 0,\ F_t\gets 900$. Hence, $l \gets (F_t - S_t)/300=3$. So, we have three schedules $\{\delta_{1},\delta_{2},\delta_{3}\}$ in which jobs will be assigned randomly for $K=5$ times.
In this example, all five iterations of random job allocations into these three schedules are shown in \cref{ujcs-tab1}. $2^{nd}$ column to $4^{th}$ column shows all 6 random job allocation into 3 different schedules. from column 5 to column 7 the total cost of each schedule is shown. Column nos. 2, 3, and 4 forms the total cost matrix. 
Each row of the cost matrix is given as input to the variance analysis part of the algorithm in which the variance of each row is calculated and compared with the variance of the next row. The row with the lowest variance is returned as the most balanced distribution of jobs. In \cref{ujcs-tab2}, the variance of each schedule after $5$ iterations is shown.
Among all the variances found in each iteration shown in \cref{ujcs-tab2}, the row with the minimum variance $(4^{th}\ row)$ is selected. The corresponding row in \cref{ujcs-tab1} is then selected as the \emph{most balanced job distribution}. The distribution of jobs of each schedule of that row is as follows: $\delta_1:<\gamma_1,\gamma_3>,\delta_2:<\gamma_2,\gamma_4,\gamma_5>,\delta_3:<\gamma_6>$
Now, in the last part of the algorithm we go a further step and select the schedule with the largest total cost first, second largest total cost next and so on.
\end{example}

\section{Analysis of the Proposed Method}\label{ana}

The following analysis are motivated by \cite{upfal2005probability} and \cite{cormen2022introduction}.
\begin{lem}{The probability of allocation of all higher cost jobs (suppose $J$) among $n$ jobs (where $|J|<n$) in a single schedule is $\frac{1}{l^J}$, where $l$ is the number of schedules.}\label{lem1}
\end{lem}
\begin{table}
 \caption{Sample Table of Schedules with their variances}\label{ujcs-tab2}
\centering
\begin{tabular}{ccccc}
\hline
\multirow{2}{*}{Iteration\#} & \multicolumn{3}{c}{Schedules}             & \multirow{2}{*}{Variance} \\ \cmidrule{2-4}
                             & $\delta_1$&$\delta_2$&$\delta3$ &                           \\ \hline
1                            & {27} & {10} & 2  & 163                       \\ 
2                            & {12} & {10} & 17 & 13                        \\ 
3                            & {13} & {15} & 11 & 4                         \\ 
4                            & {12} & {12} & 15 & 3                         \\ 
5                            & {14} & {10} & 15 & 7                         \\ \hline
\end{tabular}

\end{table}
\begin{prrf}
Let us fix any arbitrary schedule $j$ among $l$ schedules. Now, we define the event $A_{ij}$ (where $i=\{1,\cdots,J\}$) as any $i\in J$, is scheduled in an arbitrary schedule $j$. Given the event $A_{ij}$, we are interested in finding $Pr\{A_{1j}\cap A_{2j}\cdots \cap A_{Jj}\}$. If this probability is calculated then we will find that for all $i\in J$, event $A_{ij}$ will occur. So, we can write
\begin{equation}\label{eqn4}
\begin{split}
    Pr\{A_{1j}\cap A_{2j}\cap \cdots \cap A_{Jj}\}=Pr\{A_{1j}\}\cdot\\ Pr\{A_{2j}|A_{1j}\}\cdot Pr\{A_{3j}|A_{1j}\cap A_{2j}\}\cdots\\
    Pr\{A_{Jj}|A_{1j}\cap A_{2j}\cap \cdots\cap A_{(J-1)j}\}
\end{split}
\end{equation}

Now, according to our method, any higher cost jobs ($J$) may be allocated into same schedule. But allocation of one higher cost job into one schedule has no impact on the other job which is yet to be allocated into the same schedule. Consider the probability $Pr\{A_{2j}|A_{1j}\}$, where event $A_{1j}$ which has already occurred, $i.e.,$ the $(1j)^{th}$ job is already been allocated into a schedule, has no impact on the event $A_{2j}$, $i.e.,$ event $A_{2j}$ is independent of event $A_{1j}$. So we can write,
$Pr\{A_{2j}|A_{1j}\}=Pr\{A_{2j}\}$. Similarly for probability $Pr\{A_{3j}|A_{1j}\cap A_{2j}\}$ where two events $A_{1j}\ and\ A_{2j}$ which has already occurred have no impact on the event $A_{3j}$. So, it can be written as, $Pr\{A_{3j}|A_{1j}\cap A_{2j}\}=Pr\{A_{3j}\}$. In similar manner we can also write $Pr\{A_{Jj}|A_{1j}\cap A_{2j}\cap \cdots\cap A_{(J-1)j}\}=Pr\{A_{Jj}\}$.
Accordingly, we can rewrite \cref{eqn4} for the event $A_{ij}$ that job $i$ is allocated into same schedule $j$ as follows: $Pr\{A_{1j}\cap A_{2j}\cap \cdots \cap A_{Jj}\}=Pr\{A_{1j}\}\cdot Pr\{A_{2j}\}\cdots Pr\{A_{Jj}\}$.

The probability of $i^{th}$ higher cost job to be allocated into $j^{th}$ schedule is: $Pr\{A_{ij}\} = \frac{1}{l}$ So, for two higher cost jobs $Pr\{A_{ij}\cap A_{kj}\} = \frac{1}{l}\cdot\frac{1}{l},\ where, i<k$ Similarly, for $J$ higher cost jobs we have, $Pr\{A_{1j}\cap A_{2j}\cap \cdots \cap A_{Jj}\}=\frac{1}{l}\cdot\frac{1}{l}\cdots\frac{1}{l}=\left(\frac{1}{l}\right)^J=\frac{1}{l^J}$. 
\end{prrf}
\begin{observation}
When $|\Gamma|$ is large, we can say that the probability of allocating all higher cost jobs in a single schedule is much less. For example, consider 50 jobs out of 200 jobs are of higher cost, and there are 6 schedules. So the probability of allocating higher cost jobs into any one schedule out of 6 schedules is $\frac{1}{6^{50}}=1.2371931e-39$ which is very much less. It means that over repeated random allocation of jobs, the chances of accumulating higher cost jobs into any one schedule is almost 0. 
\end{observation}

\begin{lem}\label{lem2}
The expected number of jobs to be assigned in a given schedule is $\frac{|\Gamma|}{|\Delta|}$ considering an arbitrary iteration.
\end{lem}
\begin{prrf}
We have here, $|\Gamma|$ number of trials of assigning jobs into the given schedule (as $|\Gamma|$ is the total number of jobs available in our setting). Among those trials, there are successes (the sampled job is landing in the given schedule) such that each success occurs with a probability of $\mathbb{P}$ and some failure with a probability of $\mathbb{Q}=\mathbb{P}-1$. Every trial is independent and has the probability $\frac{1}{|\Delta|}$, as $|\Delta|$ is the number of schedules available. Now, we take $\mathbb{R}$ as a random variable to denote the successful assignment of jobs into the given schedule among $|\Gamma|$ trials range within $[1,\cdots,|\Gamma|]$. So the value of $\mathbb{R}$ can be represented as $\mathbb{R}=i$, where $i\in [0,\cdots,|\Gamma|]$. So, the probability of $\mathbb{R}$ for i number of successes can be written as, $Pr\{\mathbb{R}=i\}=\binom{|\Gamma|}{i}\mathbb{P}^i\mathbb{Q}^{|\Gamma|-i}$.

This is true as there are $\binom{|\Gamma|}{i}$ ways to choose $i$ number of successes among $|\Gamma|$ trials and any one such successes occurs with the probability $\mathbb{P}^i\mathbb{Q}^{(|\Gamma|-i)}$ (as in that elementary event, $i$ successes will be there along with $(|\Gamma|-i)$ number of failures). Also, we can write;

\begin{equation}\label{eqn8}
\begin{split}
    \sum\limits_{i=0}^{|\Gamma|}Pr\{\mathbb{R}=i\}=\\\sum\limits_{i=0}^{|\Gamma|}\binom{|\Gamma|}{i}\left(\frac{1}{|\Delta|}\right)^i\left(1-\frac{1}{|\Delta|}\right)^{|\Gamma|-i}=1   
\end{split}
\end{equation}
Now, the expected number of successes, $i.e.,$ expected number of jobs in every schedule can be calculated as:
\noindent \textbf{observation}
Consider the synthetic dataset where $|\Gamma|=200$, we have $\frac{|\Gamma|}{|\Delta|}=\frac{200}{4}=50$, and this value is the number of jobs allocated in each schedule. In \cref{ovbnj}, we have furnished the comparison between OffPSP and PPSJBP against the parameter number of jobs (\cref{bvo-4}) where we can see that schedules are assigned with number of jobs as:
$|\delta_1|:51, |\delta_2|:48, |\delta_3|:53, |\delta_4|:48$ which is almost balanced allocation of jobs in each of $\delta_i\in \Delta$.
The similar observations we have for our real life dataset (KDD Cup 2015 and Bus Driver Scheduling) in the simulation (given in \cref{bvo-5} and \cref{bvo-6}).
\begin{equation}\label{eqn11}
\resizebox{\columnwidth}{!}{$%
\begin{split}
        E[\mathbb{R}]&=\sum\limits_{i=0}^{|\Gamma|} i\cdot Pr\{\mathbb{R}=i\}\\
        &=\sum\limits_{i=1}^{|\Gamma|} i\binom{|\Gamma|}{i}\left(\frac{1}{|\Delta|}\right)^i\left(1-\frac{1}{|\Delta|}\right)^{|\Gamma|-i}\\
        &=\sum\limits_{i=1}^{|\Gamma|}i\left(\frac{|\Gamma|!}{i!(|\Gamma|-i)!}\right)\left(\frac{1}{|\Delta|}\right)^i\left(1-\frac{1}{|\Delta|}\right)^{|\Gamma|-i}\\
        &=\sum\limits_{i=1}^{|\Gamma|}i\left(\frac{|\Gamma|(|\Gamma|-1)!}{i(i-1)!(|\Gamma|-i)!}\right)\left(\frac{1}{|\Delta|}\right)^i
        \left(1-\frac{1}{|\Delta|}\right)^{|\Gamma|-i}\\
        &=\sum\limits_{i=1}^{|\Gamma|}\left(\frac{|\Gamma|(|\Gamma|-1)!}{(i-1)!(|\Gamma|-i)!}\right)\left(\frac{1}{|\Delta|}\right)^i\left(1-\frac{1}{|\Delta|}\right)^{|\Gamma|-i}\\
        &=\frac{|\Gamma|}{|\Delta|}\sum\limits_{i=1}^{|\Gamma|}\binom{|\Gamma|-1}{i-1}\left(\frac{1}{|\Delta|}\right)^{i-1}\left(1-\frac{1}{|\Delta|}\right)^{|\Gamma|-i}\\
        &\ \ \ \ \ \ \ \ \ \ \ \ \ \ \ \ \ \ \ \ \ \ \ [as\ (|\Gamma|-1-i+1)!=(|\Gamma|-i)!]\\
        &=\frac{|\Gamma|}{|\Delta|}\sum\limits_{i=0}^{|\Gamma|-1}\binom{|\Gamma|-1}{i}\left(\frac{1}{|\Delta|}\right)^{i}\left(1-\frac{1}{|\Delta|}\right)^{(|\Gamma|-1)-i} \\
        &=\frac{|\Gamma|}{|\Delta|} \ \ \ \ \ \ \ \ [by\ \cref{eqn8}]\\
\end{split}$%
}
\end{equation}
\end{prrf}
From this proof, it is evident that our proposed PPSJBP distributes the jobs in each schedule in a balanced way.

\noindent \textbf{observation}
Consider the synthetic dataset where $|\Gamma|=200$, we have $\frac{|\Gamma|}{|\Delta|}=\frac{200}{4}=50$, and this value is the number of jobs allocated in each schedule. In \cref{ovbnj}, we have furnished the comparison between OffPSP and PPSJBP against the parameter number of jobs (\cref{bvo-4}) where we can see that schedules are assigned with number of jobs as:
$|\delta_1|:51, |\delta_2|:48, |\delta_3|:53, |\delta_4|:48$ which is almost balanced allocation of jobs in each of $\delta_i\in \Delta$.
The similar observations we have for our real life dataset (KDD Cup 2015 and Bus Driver Scheduling) in the simulation (given in \cref{bvo-5} and \cref{bvo-6}).

\begin{lem}\label{lem3}
The expected number of jobs to be sampled so that every schedule is assigned with $|\mathbb{Z}|\geq1$ jobs is approximately  $|\Delta|\ln |\Delta|$, where $|\mathbb{Z}|$ is the number of assigned jobs in a schedule.
\end{lem}
\begin{prrf}
Let us consider the number of jobs that have to be sampled before every schedule gets a job is $\mathbb{M}^*$ and the allocation of a job into an empty schedule a "success." In this proof, we will apply the concept of geometric distribution as we know that the number of jobs to be sampled before every schedule gets atleast one job follows geometric distribution. We shall divide the sampling of jobs into partitions. Each partition will then boil down to the geometric distribution. The $i^{th}$ success requires sampling of jobs after the $(i-1)^{st}$ successes.

In the first partition, since all the schedules are initially empty, a schedule will get a job with probability $\frac{|\Delta|}{\Delta}=1$. 
When one schedule has got one job, sampling of jobs in second partition may allocated an arbitrary job either into the non-empty schedule or into any one of the empty schedules. As there are $(|\Delta|-1)$ empty schedules, while sampling in second partition, a job will be assigned in one of the $(|\Delta|-1)$ empty schedules with the probability $\frac{(|\Delta|-1)}{|\Delta|}$. 
In $i^{th}$ partition, there are $(|\Delta|-i+1)$ number of empty schedules as $(i-1)$ schedules are allocated with atleast one job in $(i-1)$ partitions. Hence, the probability of getting a job into an empty schedule in $i^{th}$ partition is $\frac{(|\Delta|-i+1)}{|\Delta|}$.
In each partition the number of sampling of jobs can be represented with a random variable $\mathbb{M}^*_i$ where $i$ lies within the range of $\{1,\cdots,|\Delta|\}$. So, in order to get $|\Delta|$ successes we need $\mathbb{M}^*=\sum\limits_i^{|\Delta|}\mathbb{M}^*_i$ number of samplings. As in each partition, the random variable $\mathbb{M}^*_i$ follows geometric distribution and probability of success is $Pr\{\mathbb{M}_i^*\}=\frac{(|\Delta|-i+1)}{|\Delta|}$ So the expected number of samples in $i^{th}$ partition is thus $E[\mathbb{M}^*_i]=\frac{|\Delta|}{|\Delta|-i+1}$

Now the expected number of sampling of jobs required in all partitions can be obtained as:
\begin{equation*}
\begin{split}
    E[\mathbb{M}^*]&=E\left[\sum\limits_{i=1}^{|\Delta|} \mathbb{M}^*_i\right]\\
                 &=\sum\limits_{i=1}^{|\Delta|} E[\mathbb{M}^*_i]\ \ \ \ \ by\ linearity\ of\ expectation\\
                 &=\sum\limits_{i=1}^{|\Delta|} \frac{|\Delta|}{|\Delta|-i+1}\ \ \ \ \ \ by\  E[\mathbb{M}^*_i]\\
                 &=|\Delta|\sum\limits_{i=1}^{|\Delta|} \frac{1}{i}\\
                 &=|\Delta|(\ln|\Delta|+\mathcal{O}(1))\ \ \ \ \ by\ Harmonic\ series
\end{split}
\end{equation*}
\end{prrf}

\begin{lem}\label{lem4}
The probability $Pr\{\mathbb{R}\geq 1.5\mathbb{E}[\mathbb{R}]\}\le e^{-\frac{|\Gamma|}{12|\Delta|}}$, for any schedule $\delta_i,$ where $i\in [1,|\Delta|]$.
\end{lem}
\begin{prrf}
We can find the probability $Pr\{\mathbb{R}\geq 1.5\mathbb{E}[\mathbb{R}]\}$ using Chernoff Bound as: $Pr(\mathbb{R}\geq(1.5)\mathbb{E}[\mathbb{R}])=Pr(\mathbb{R}\geq(1+\frac{1}{2})\mathbb{E}[\mathbb{R}])\le e^\frac{-\mathbb{E}[\mathbb{R}]{(\frac{1}{2})}^2}{3}$.

We know that $\mathbb{E}[\mathbb{R}]=\frac{|\Gamma|}{|\Delta|}$. So, plugging this in the above equation we get:

\begin{equation*}
\begin{split}
    Pr(\mathbb{R}\geq(1.5)\mathbb{E}[\mathbb{R}])&\le e^{-\frac{\mathbb{E}[\mathbb{R}]{(\frac{1}{2})}^2}{3}}\\
    &\le e^{-\frac{|\Gamma|}{3|\Delta|}\cdot (\frac{1}{2})^2}\\
    &=e^{-\frac{1}{4}\cdot\frac{|\Gamma|}{3|\Delta|}}\\
    &=e^{-\frac{|\Gamma|}{12|\Delta|}}
\end{split}
\end{equation*}
\end{prrf}
From the above discussion we can see that a little deviation of any schedule from balanced job assignment has the probability is upper bounded by $e^{-\frac{|\Gamma|}{12|\Delta|}}$. Now, as $|\Gamma|$ grows larger, $Pr{(\mathbb{R}\geq(1.5)\mathbb{E}[\mathbb{R}])}$ will be decreasing exponentially.
\begin{observation}
With $|\Gamma|=200$ and $\Delta=4$, the above said probability can be calculated as $e^{-\frac{200}{12\times4}}=e^{-4.17}=0.0154$, which is much less. In terms of our dataset, namely, Bus Driver Scheduling dataset and KDD Cup 2015 dataset, $|\Gamma|=359$ and $|\Gamma|=3000$ respectively. So the said probability will be much less than the case where $|\Gamma|=200$.
\end{observation}
\begin{lem}\label{lem5}
The time complexity by PPSJBP is $\mathcal{O}(K\times|\Gamma|)+\mathcal{O}(|\Delta|\log |\Delta|)$, where $K$ is the number of repetition of sampling of jobs, $|\Gamma|$ is the total number of jobs and $|\Delta|$ is the number of schedules.
\end{lem}
\begin{prrf}
Our algorithm PPSJBP consists of three subroutines - \emph{RRA}, \emph{MBDF}, and \emph{LCSF}. In \emph{RRA} $|\Gamma|$ number of jobs are allocated randomly in $|\Delta|$ schedules and this random allocation is repeated for $K$ times. That means for each of the $K$ iterations, $|\Gamma|$ of jobs have been allocated into $|\Delta|$ schedules. So the time complexity for RRA subroutine is $\mathcal{O}(K\times |\Gamma|)=\mathcal{O}(K|\Gamma|)$. MBDF subroutine finds the most balanced set of schedules by calculating the variance of each set of schedule out of $K$ such sets. Variance calculation of each set takes $\mathcal{O}(|\Delta|)$ time. This is for one iteration. So, variance calculation for $K$ such set of schedules take $\mathcal{O}(K\times |\Delta|),\ i.e., \mathcal{O}(K|\Delta|)$ time. 
Now, in the last subroutine (LCSF), the minimum variant set of schedules i.e., most balanced set of schedules are arranged in descending order according to their costs. To achieve that arrangement, merge sort is used. The input for the sorting is $|\Delta|$ number of costs (as each schedule will provide one cost). So the time complexity for LCSF subroutine is $\mathcal{O}(|\Delta|\log |\Delta|)$.
Accordingly, the total time complexity for the three subroutine can be calculated as $\mathcal{O}(K|\Gamma|)+\mathcal{O}(K|\Delta|)+\mathcal{O}(|\Delta|\log |\Delta|)=\mathcal{O}(K|\Gamma|)+\mathcal{O}(|\Delta|\log |\Delta|)$. If $|\Delta|<<|\Gamma|$ (which is most realistic scenario), the time complexity of PPSJBP is $\mathcal{O}(K|\Gamma|)$.
\end{prrf}

Now a connection will be established between finding our minimum variance with repetition and online hiring assistant problem for stating our next analytical result. In online hiring assistant problem, there are $m$ number of candidates who appear for the interview process in an online fashion $i.e.,$ we don't know the list of the candidates apriori.
In this problem we don't want to interview all the $m$ number of candidates yet, we want to find the best candidate or a close to the best candidate (that is the target). The natural question that arises is that, how many samples we need to explore before we are sure that our target is achieved with a good probability. In literature \cite{cormen2022introduction}\cite{ajtai2001improved}\cite{hajiaghayi2004adaptive}\cite{kleinberg2005multiple}\cite{bateni2013submodular}\cite{correa2020sample}\cite{nuti2022secretary}, it is shown that we need to sample $\frac{m}{e}$ number of candidates to make sure that our target is achieved with a probabilistic lower bound of $\frac{1}{e}$.
In our problem, for each iteration, we calculate a variance. So, as a thought experiment, we can think the variances are appearing in an online fashion just like the candidates in the online hiring assistant problem. So the variances here are resembling the candidates and we need to find out the best (minimum) variance. So, the problem is boiling down to the fact that how many such variances to be sampled (to be calculated) before we achieve our target of finding the minimum variance.
\begin{corr}\label{cr1}
$\frac{K}{e}$ number of variances to be calculated (per iteration we calculate one variance) to make sure that the minimum variance will be achieved with a probabilistic lower bound $\frac{1}{e}$.
\end{corr}
\begin{prrf}
The proof follows from the connection we have established above, $i.e.,$ we need $\frac{m}{e}$ number of samples for the probability to be atleast $\frac{1}{e}$. Following this connection, we can say that $\frac{K}{e}$ number of variances to be calculated for making sure that we get a probabilistic lower bound of $\frac{1}{e}$ ($i.e,$ with probability atleast $\frac{1}{e}$).
\end{prrf}
\begin{observation}
Consider the value of $K=4000$ repeated random allocation of jobs. According to Corollary \ref{cr1}, we have to iterate the random allocation of jobs into schedules $\frac{K}{e}$ times $i.e.,$ $\frac{4000}{e}\approx 1472$ times to get success with the probability of $1/e\approx0.37$ $i.e.,$ if we the repeat random allocation process for 1472 times, the chance to get best minimum variance is atleast 37\%. Now, if we iterate our random allocation process for $K$ times, the lower bound of the probability will also increase. So, we can say that with larger value of $K$, the lower bound of the probability of getting the best minimum variance is greater than $\frac{1}{e}$.
\end{observation}

\section{Experiments and Results}\label{er}
\subsection{\textbf{Setup}}\label{st}
\noindent \textbf{Number of Schedules:} 
It is stated earlier in Section \ref{moddef} that in our model a time horizon $T$ is divided into a set of schedules $\Delta=\{\delta_1,\delta_2,\cdots,\delta_l\}$ and the time duration of each $\delta_i$ is $\tau_i$ with $\tau_1=\tau_2=\cdots=\tau_{|\Delta|}$. So $|\Delta|$ was defined as $|\Delta|=\frac{T}{\tau_i},\ where\ \tau_i\in \mathbb{R}, \tau_i\ne0,\ and\ \tau_i\le T$. It is to be noted that, $\tau_i$ depends on the applications we are working in and thus our algorithm is flexible enough to be amenable to any such application. 
In our algorithm, the number schedules ($|\Delta|$) taken is 4 and the total duration ($T$) for all tasks is considered 4 weeks for simulation. So $\tau_i=1$ week $\forall_i$.

\noindent\textbf{Dataset:} Another vital part of the setup is the dataset. First, a synthetic dataset was generated with random numbers as job cost for primary simulation of our algorithm. It was generated using Python Libraries. After the successful simulation with the synthetic dataset, we have used two real datasets, - i) Bus Driver Scheduling dataset \cite{constantino2017solving} and ii) KDD Cup 2015 dataset \cite{feng2019understanding}. All datasets are elaborated in \cref{ds}.

\noindent\textbf{System Specification:} The simulation is carried out in personal computing environment with - $5^{th}$ generation microprocessor, 8 GB of RAM, Ubuntu as the operating system. Google Colab (with Python) is used for implementing the algorithm. 

\subsection{Dataset}\label{ds}
\noindent\textbf{Synthetic Dataset:} The synthetic dataset, shown in \cref{ujcs-tab3} consists of the columns- $job\  label$ ($i.e., (\gamma_i \in \Gamma)$) and $job\ cost$ ($\chi \in \mathbb{Z}^+$). This dataset has been generated randomly with 200 jobs. Once we have verified our algorithm with this synthetic dataset, we have employed \text{two real datasets} - Bus Driver Scheduling dataset \cite{constantino2017solving} and KDD Cup 2015 dataset \cite{feng2019understanding} in our algorithm.

\begin{table}
\caption{Sample of Synthetic Dataset}\label{ujcs-tab3}
\centering
 \begin{tabular}{@{}ccc@{}}
\toprule
 Sl \# & Job ($\Gamma$) & Job Cost ($\chi)$ \\ 
 \midrule
 1 & $\gamma_1$ & 10 \\ 
 2 & $\gamma_2$ & 40 \\
 3 & $\gamma_3$ & 25 \\
 4 & $\gamma_4$ & 100 \\
 5 & $\gamma_5$ & 89 \\
 \hline
 \end{tabular}
\end{table}

\begin{table}
\caption{Total Cost}\label{b}
\centering
\begin{tabular}{@{}ccc@{}}
\toprule
Sl\#& Bus\_ID & Tot\_Dur\\
\midrule
1&1&1145\\
2&2&1110\\
3&3&850\\
4&4&1090\\
5&5&840\\
\hline
\end{tabular}
\end{table}

\noindent\textbf{Bus Driver Scheduling Dataset:} Bus Driver Scheduling dataset has seven columns or attributes all of which are not required for the algorithm. As we are only concerned about the job\_id and the job cost, we have extracted the values associated with rows "Bus\_Line\_Id" and "Duration". In this dataset there are a total of 359 buses that run per day. 
The total running cost of every bus for the entire day is obtained by adding all the run time duration of that day. For example, $Bus\_Line\_Id: 1$ total running time is 1145 minutes. The 5 samples of the dataset after summing up the run time cost of every bus is shown in \cref{b}.
\begin{table}
\caption{Partial Extraction from Course Log Information}\label{logex}
\begin{tabular}{ccccc}
\toprule
enroll\_id     & Date       & Time       & event       &course\_id \\ 
\midrule
76492          & 2013-12-12 & 03:46:04   & navigate    & 81UZ       \\ 
115995         & 2013-11-28 & 12:00:10   & video       & 3VkH       \\ 
124335         & 2014-05-13 & 09:37:03   & access      & q6A6       \\ 
81407          & 2014-01-01 & 08:03:01   & access      & 5Gyp       \\ 
81407          & 2014-01-01 & 08:03:01   & video       & 5Gyp       \\ 
\hline
\end{tabular}
\end{table}

\begin{table}
\caption{Extraction of Event Cost Per Course}\label{jbcst}
\centering
\begin{tabular}{ccc}
\hline
course\_id            & Job         & Cost \\ \hline
\multirow{3}{*}{81UZ} & navigate    & 10   \\  
                      & access      & 15   \\  
                      & video       & 4    \\ \hline
\multirow{3}{*}{G8EP} & video       & 1    \\  
                      & page\_close & 1    \\  
                      & discussion  & 4    \\ \hline
\multirow{4}{*}{3cnZ} & access      & 2    \\  
                      & video       & 1    \\  
                      & page\_close & 1    \\  
                      & discussion  & 5    \\ \hline
\end{tabular}

\end{table}
\noindent\textbf{KDD Cup 2015 Dataset:} KDD Cup 2015 dataset is a very large repository which contains online course enrollment and course log information of a large number of students. The interpretation of this dataset is as follows: a student can enroll in multiple courses, and every course has at most $7$  activities or events, which are - \emph{Access}, \emph{Video}, \emph{Discussion}, \text{Navigate}, \emph{Problem}, \emph{Wikipedia}, \emph{Page  Close}. Each activity under every course in which a particular student has enrolled, is considered as a job in this paper.  
Each course log information entry has the information about a student's - enrollment\_id, entry time, activity name, and course\_id. We have extracted the log information of 30 students who has enrolled for maximum courses.
From course log information entry, the column $"time"$ is divided into $"date"$ and $"time"$ and put into a new entry (shown in \cref{logex}). Then the total activity time for each event under each course is calculated by taking the difference between the $"entry\_time"$ of next event and the $"entry\_time"$ of that event. These differences of two different entry times are then summed up for each event under each course which is shown in Table \cref{jbcst}. Again, for the course\_id column, first four characters are shown. The total cost of an event under a course (considered as the job) is given in minutes. There are 3000 such jobs in our derived dataset, sample of which is presented in \cref{jbcst}. The implementation of extraction of both the real datasets is done using Python with Pandas library.

\subsection{Results of Simulation}\label{rs}
In this section we have presented the outcome of simulation of our algorithm on various parameters.

\subsubsection{Comparison Among the Set of Schedules with Minimum Variance and Randomly Chosen Sets of Schedules}\label{seslf-cmp1}
This section shows the comparison between the most balanced set of schedules and the other randomly chosen sets of schedules which are unbalanced. Let us represent the most balanced set of schedules as $\Delta^{Balanced}$ and randomly chosen unbalanced set of schedules as $\Delta^{Other}$. The setup for this simulation is given below. 

\[\min_{k}\{Var(\Delta_k)\}\ and\ j= \arg \min_{k}{Var(\Delta_k)}\]
\[\Delta^{\prime}=\{rand(\Delta_{k\neq j})\},\ where\ |\Delta^{\prime}|=k^{'}<K\]
\resizebox{\columnwidth}{!}{$\Delta^{\prime\prime}=\{rand(\Delta_{k\neq j})\ and\ \Delta_k\notin\{\Delta^{'}\}\},\ where\ |\Delta^{\prime\prime}|=k^{'}<K$}

\begin{figure}
\centering
\begin{subfloat}[Comparison Using Synthetic Dataset]{
    \includegraphics[width=0.8\linewidth]{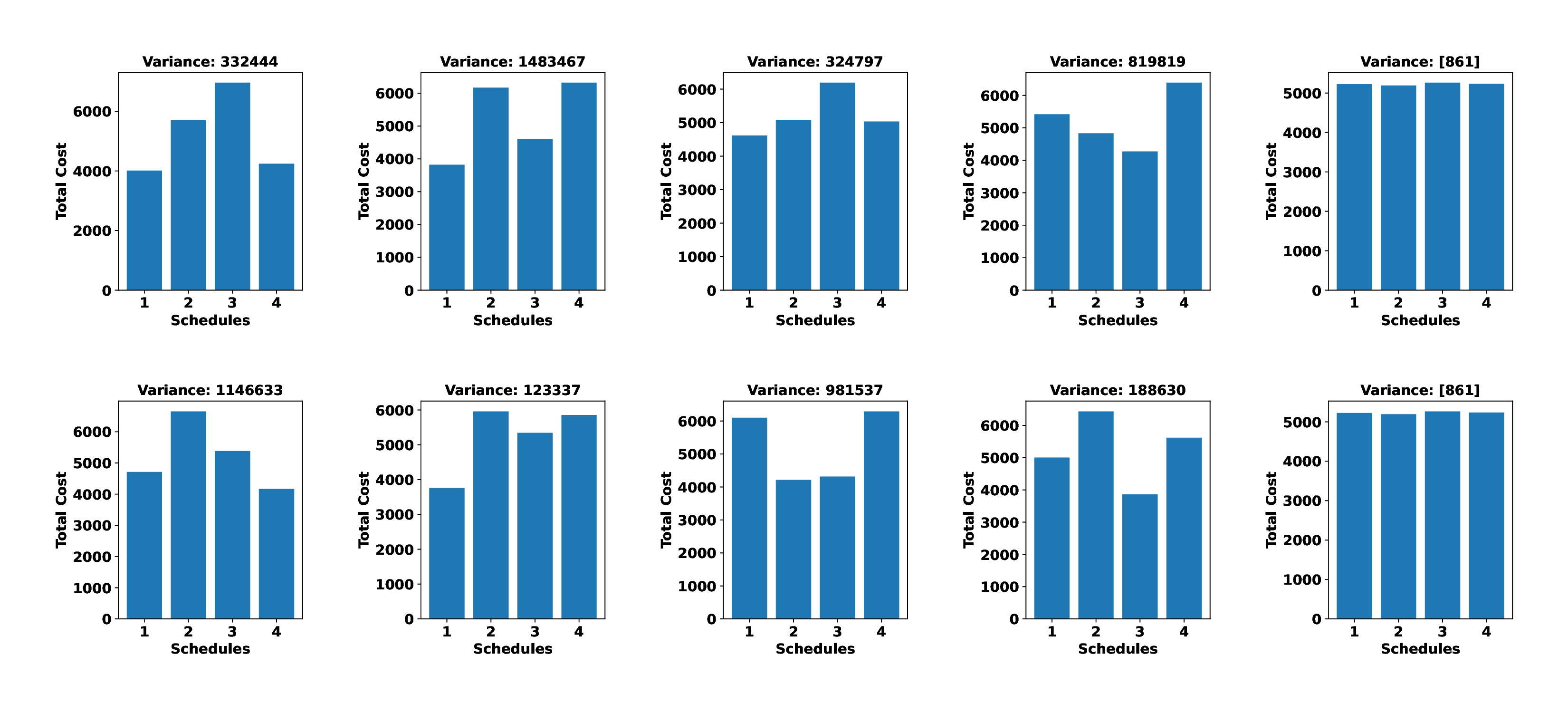}
    \label{er-fig-1}}
  \end{subfloat}
  \begin{subfloat}[Comparison Using Bus Driver Scheduling Dataset]
  {
    \includegraphics[width=0.8\linewidth]{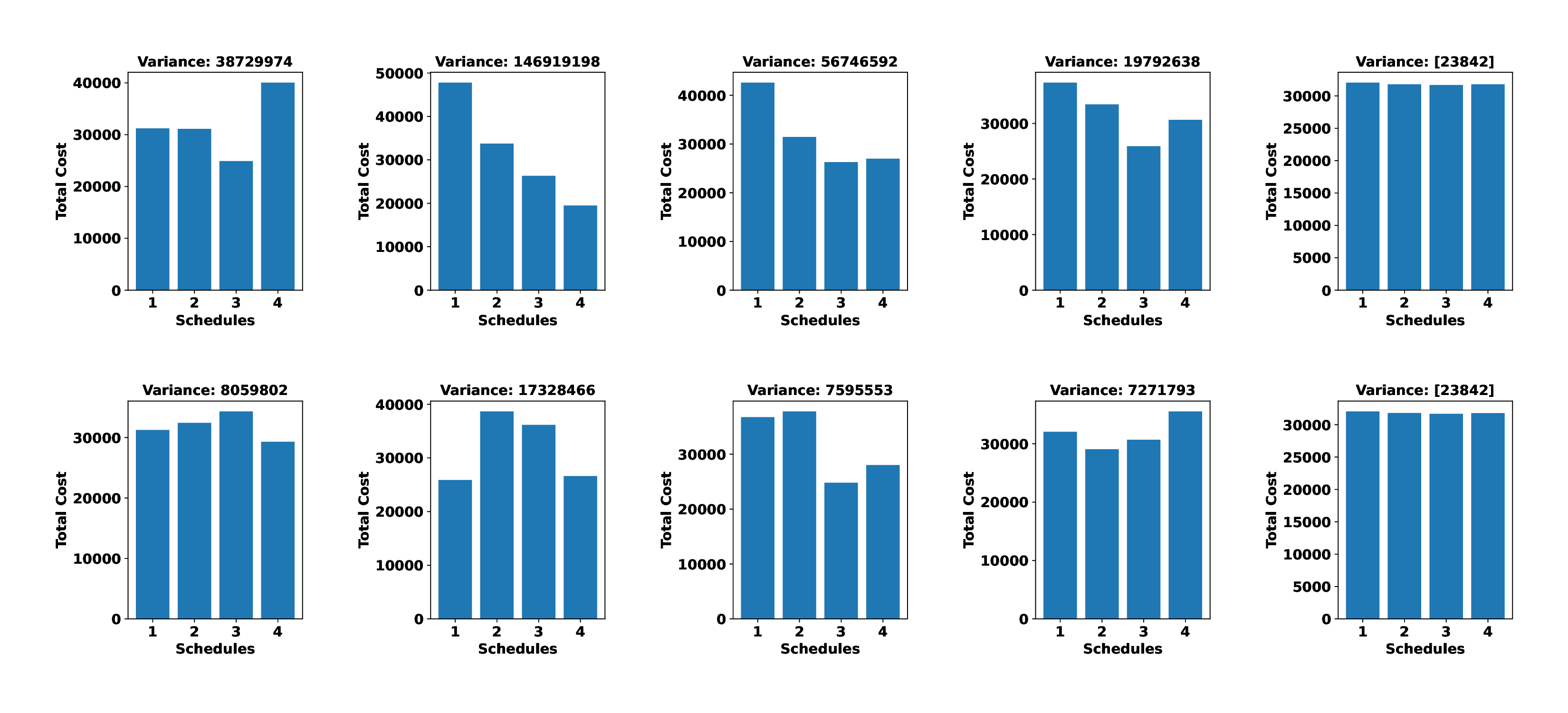}
    \label{er-fig-2}}
  \end{subfloat}
  \caption{\centering Comparison Between Randomly Chosen Set of Schedules With the Minimum Variance set} \label{BU_Com}
\end{figure}

In this setting, $j$ is the index position of the set of schedules with minimum variance. $\Delta^{\prime}$ is the $k^{'}$ sets of schedules taken randomly and $\Delta^{\prime\prime}$ is the another $k^{'}$ sets of schedule taken randomly except the first $k^{'}$ sets. $rand(.)$ is the function which picks the $k^{th}$ set of schedules at random. For our case $k^{'}=4$. This comparison is shown using synthetic dataset with $K=8000$, shown in \cref{er-fig-1} and then using the Bus Driver Scheduling dataset with $K=25000$, shown in \cref{er-fig-2}. In both the figures, first row of the figure contains five plots among which first four plots are from set of $\Delta^{Other}$ and the last plot is of $\Delta^{Balanced}$. The second row is same but without replacement by the first four of $\Delta^{Other}$ and the fifth plot is of $\Delta^{Balanced}$ again. The purpose of this comparison is to show that set of schedules with minimum variance is most balanced one in comparison to any other sets of schedules obtained from repeated random allocation. When the number of jobs increases, the repetition of random allocation is also to be increased in order to lower the probability of concentration of higher cost jobs into one schedule (refer to Lemma \ref{lem1}).
In \cref{er-fig-1}, it is to be noticed that the difference of variances of first four $\Delta^{Other}$ and $\Delta^{Balanced}$ is very large, $e.g.,$ first four schedules have variances of $332444$, $14834678$, $324797$, $819819$ and the most balanced one has the variance of $861$ (shown in first row of \cref{er-fig-1}). It is evident from the above 
simulation that our algorithm is able to allocate all jobs into schedules in most balanced manner according to their costs. 
This simulation follows Lemma \ref{lem1} which states that the accumulation of all higher cost jobs into a single schedule is very less. We claim this because when the minimum variant schedules is found after $K$ iterations, we have seen that the cost of each schedule is almost equal. 

The connection we have established in \cref{ana} reveals that the number of variances to be calculated to achieve the best minimum variance with a lower bound probability of 0.37, is approximately $1472$ (when $K=4000$). That means if we calculate $1472$ number of variances ($i.e.,$ the number of iterations), there is atleast $37\%$ chance to obtain the best minimum variance. When we take $K=8000$, to achieve the lower bound of 0.37 we need to find $2943$ number of variances. In order to increase this lower bound, the value of $K$ is to be increased because this lower bound of $37\%$ is giving a benchmark. If we compute less number of variances, the scenario would not be realistic. So increasing value of $K$ will give higher accuracy. Accordingly, in our simulation we have considered $8000$ and more number of iterations which is sufficiently larger than the value given in the benchmark.  

From Lemma \ref{lem3} we can say that the expected number of jobs to be sampled is $|\Delta|\ln|\Delta|$. That means, if we sample $|\Delta|\ln|\Delta|$ jobs, all the schedule will be assigned with one or more jobs. From this simulation (in \cref{BU_Com}), we have seen that in the minimum variant set of schedules all the schedules have been assigned with jobs with almost balanced distribution. Also, in the randomly chosen sets, we have seen that all the schedules are assigned with one or more jobs and no schedule is left empty. Thus, the observation made from this simulation is plausible enough to justify Lemma \ref{lem3}. 

\subsubsection{Comparative Study Varying \emph{K}}\label{csv} 
\begin{figure}
\centering
    \includegraphics[width=0.8\linewidth]{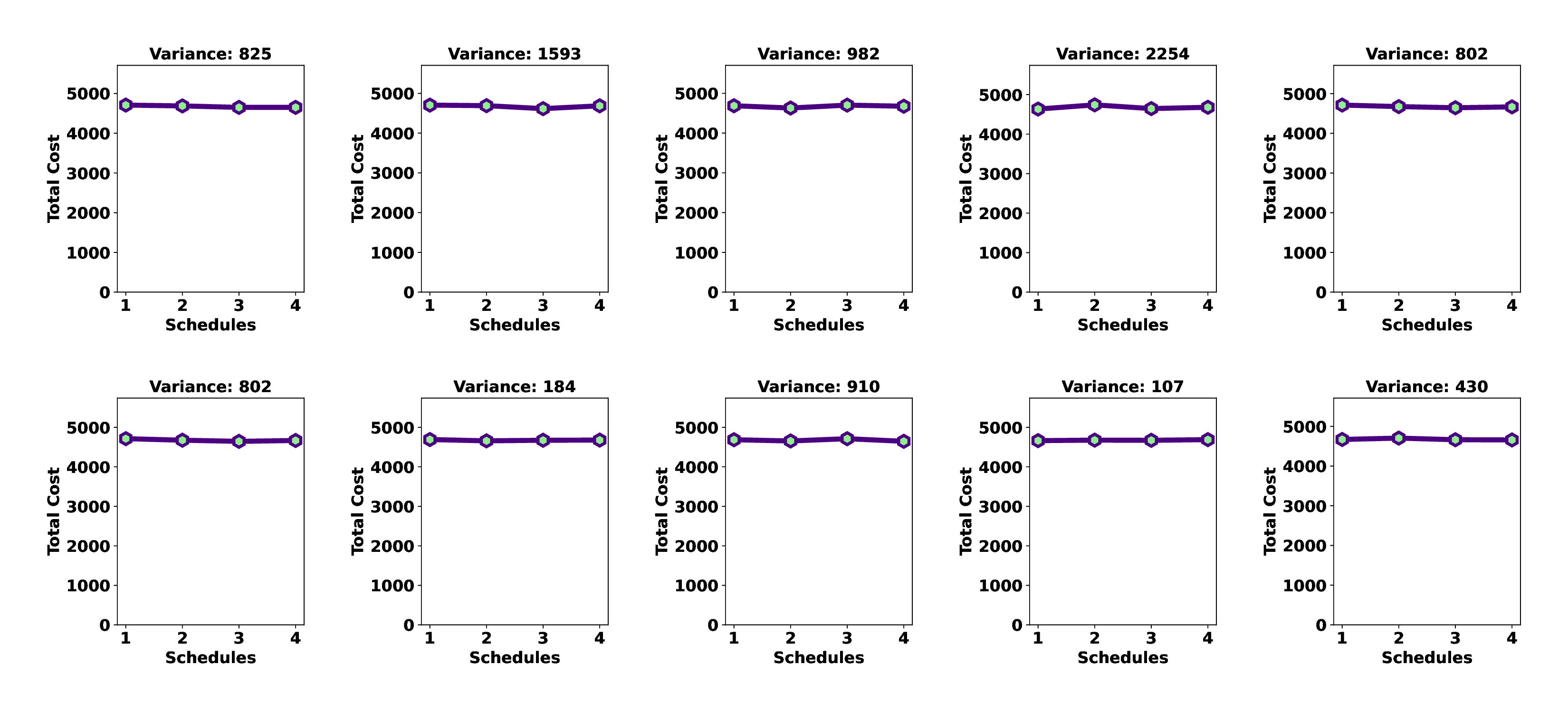}
    \caption{Comparison of Balanced Sets of Schedules over Multiple $K$ Iterations}
    \label{er-fig-3}
\end{figure}

In this simulation we have shown a comparative study on the behaviour of $\min_{k}\{Var(\Delta_k)\}$ every time when $K$ is varied. Here, we have varied the value of $K$ from $10000$ to $20000$ and in each case it has been observed that our algorithm is able to obtain the most balanced set of schedules. The comparison is given in \cref{er-fig-3}. From this simulation, we can claim that the proposed mechanism will always produce the most balanced set of schedules which ensures the most uniform distribution of jobs in each set of schedules.

In each plot, $X$ axis denotes the schedule number and $Y$ axis denotes the total cost of each schedule (\cref{er-fig-3}).

Lemma \ref{lem4} shows that our algorithm has a negligible chance for any schedule to deviate from getting the balanced job assignment. This simulation shows that in each of multiple $K$ iterations, the minimum variant set of schedules is achieved and it is balanced.
\subsubsection{Comparative Study on Dispensing of Schedules with and without Ordering}
\begin{figure}
\centering
\begin{subfloat}[b][Comparison Using Synthetic Dataset]{
    \includegraphics[width=0.45\linewidth]{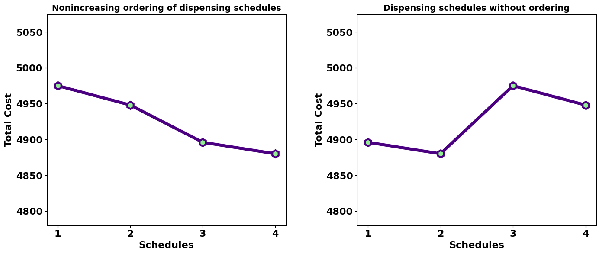}
    \label{sort-fig-1}}
  \end{subfloat}
  \begin{subfloat}[b][\centering Comparison Using Bus Driver Scheduling Dataset]
  {
   \includegraphics[width=0.45\linewidth]{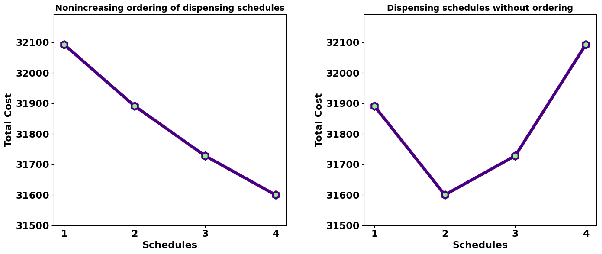}
   \label{sort-fig-2}}
  \end{subfloat}
  \begin{subfloat}[b][\centering Comparison Using KDD Cup 2015 Dataset]
  {
   \includegraphics[width=0.45\linewidth]{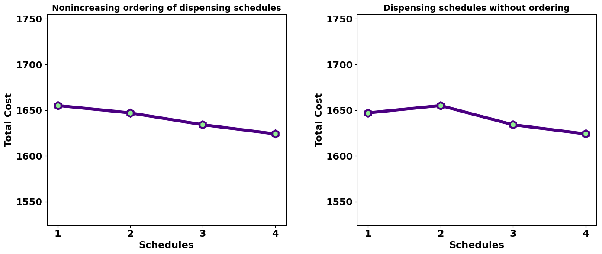}
   \label{sort-fig-3}}
  \end{subfloat}
  \caption{Comparison With and Without Ordering Schedules}
\end{figure}

The last step of our proposed method is to dispense the schedules in nonincreasing order of their costs. This step is followed after we have found the minimum variant set of schedules. In the minimum variant set, the total cost of each is different but those costs are in close vicinity (refer \cref{seslf-cmp1} and \cref{csv}). From that set, the schedule with the maximum cost is dispensed first, then the schedule with second maximum cost and so on. The representation of such ordering is given below.
\[\Delta^{\prime\prime\prime}=Sort(\min_{k}\{Var(\Delta_k)\})\] where $Sort()$ function rearranges the set of schedules with minimum variance in nonincreasing order of their cost. 
Through such ordering of dispensing of schedules, the agent will have maximum time to verify the jobs of heavier schedules after submission. In this method of dispensing of schedules, the submission and verification process are carried out simultaneously and the verification of all jobs could be performed efficiently. In \cref{sort-fig-1}, the simulation in left hand side shows the dispensing of schedules with nonincreasing order of costs and right hand side shows without ordering. In the right hand side of \cref{sort-fig-1}, we can see that if schedule 2 is dispensed before schedule 3, the verification time for schedule 3 would be less than schedule 2 even if schedule 3 is heavier than schedule 2. In the same figure, the left hand side shows that the schedule with largest cost, $i.e.,$ schedule 3 is dispensed first, then schedule 4 and so on which provides the maximum time for each schedule for verification after submission of jobs in those schedules. 

In this simulation the minor difference among the cost of each schedule is amplified so that the dispense of schedules with and without ordering could clearly be understood.

\section{Comparison with Existing Research}\label{cmpr}
Our proposed scheduling mechanism that subvert the procrastinating nature of the agents is compared with the existing work in \cite{wang2019procrastination}. Through this comparison, we will discuss that our algorithm (PPSJBP) will generate a set of schedules which is much more balanced than the algorithm (OffPSP) mentioned in \cite{wang2019procrastination}.

According to the algorithm OffPSP, a threshold regarding the cost of each schedule is fixed and all the jobs are allocated according to threshold to every schedule. In our algorithm PPSJBP, we have made a balanced allocation of all jobs into each schedule without taking any threshold (this gives a general flavour to our proposed algorithm as threshold based allocation is somewhat imposing extra restriction on the allocation of jobs to the schedule). In OffPSP the threshold based technique leads to a set of schedules where every schedule in that set is unbalanced (in terms of its costs).

Every schedule in OffPSP has a time period $t\in T$ (T is the set of time periods). Each job $j\in J$ ($J$ is the set of jobs) that is assigned to a particular schedule has a start time and a deadline such that the job can be completed within the time period of that schedule. That means, the time period of a schedule is larger than the time period of a job assigned to that schedule.
The job set and the time frame (of a schedule) in which a job can be completed, is mapped to a bipartite graph, from which each job is allocated to a schedule after satisfying a condition of maximum ratio of marginal utility to the cost. In this algorithm a threshold is maintained (already mentioned) which is violated once by the last allocated job of the current schedule and then for that schedule, the utility of the last job is checked with that of the other jobs in that schedule. Job with the higher utility is kept and the rest are discarded from the schedule.

We have implemented the OffPSP algorithm in our setting where we have considered that all jobs have uniform utility (this is taken as uniform to have the equivalence of the setting of OffPSP and our proposed algorithm PPSJBP). According to the OffPSP algorithm, a job is allocated to a schedule if the the ratio of marginal utility to the cost of that job is maximum compared to all other jobs left in that schedule. As we have considered the uniform utility for all jobs, the ratio is calculated as $\frac{utility}{cost}$. Jobs are allocated into a particular schedule until the threshold (cost) of that schedule is violated once (the threshold is defined in \cref{ovp}).
Let us understand the procedure with an example. Consider, the job set is $J={(j_1,12),(j_2,50),(j_3,4),(j_4,45),(j_5,16),(j_6,9)}$, number of schedules are $2$, and the threshold for each schedule is $55$. The ratio of marginal utility and cost for each job is calculated as $\frac{1}{j_i},\ where\ j_i\in J$, as the marginal utility of the schedule after newly added job to that schedule and before adding it, is $1$ (definition of marginal utility is given in \cite{wang2019procrastination}). Now, after calculating the ratio, we found that the $3^{rd}$ job with cost $4$ has the maximum ratio marginal utility to the cost. So it is allocated to schedule 1 as the total cost of schedule 1 is less than the threshold. The allocated job $j_3$ is them removed from the job pool. The total cost of that schedule becomes 4. In this way, the jobs $j_3,\ j_6,\ j_1,\ j_5,\ and\ j_4$ are assigned to schedule 1. The total cost for schedule 1 becomes 86. Now the total cost is more than the threshold. As the condition is violated once the allocation process into schedule 1 is stopped. Now the only job left is $j_2$ which has the cost 45. This job is allocated to schedule 2 as it is empty. So, finally in schedule 1 and schedule 2, the total job cost is 86 and 45, and number of jobs are 5 and 1 respectively. From this example we can see that both the schedules produced, are not balanced in terms of the total cost and the number of jobs. 
\begin{figure}
\centering
\begin{subfloat}[b][Synthetic]{
    \includegraphics[width=0.30\linewidth]{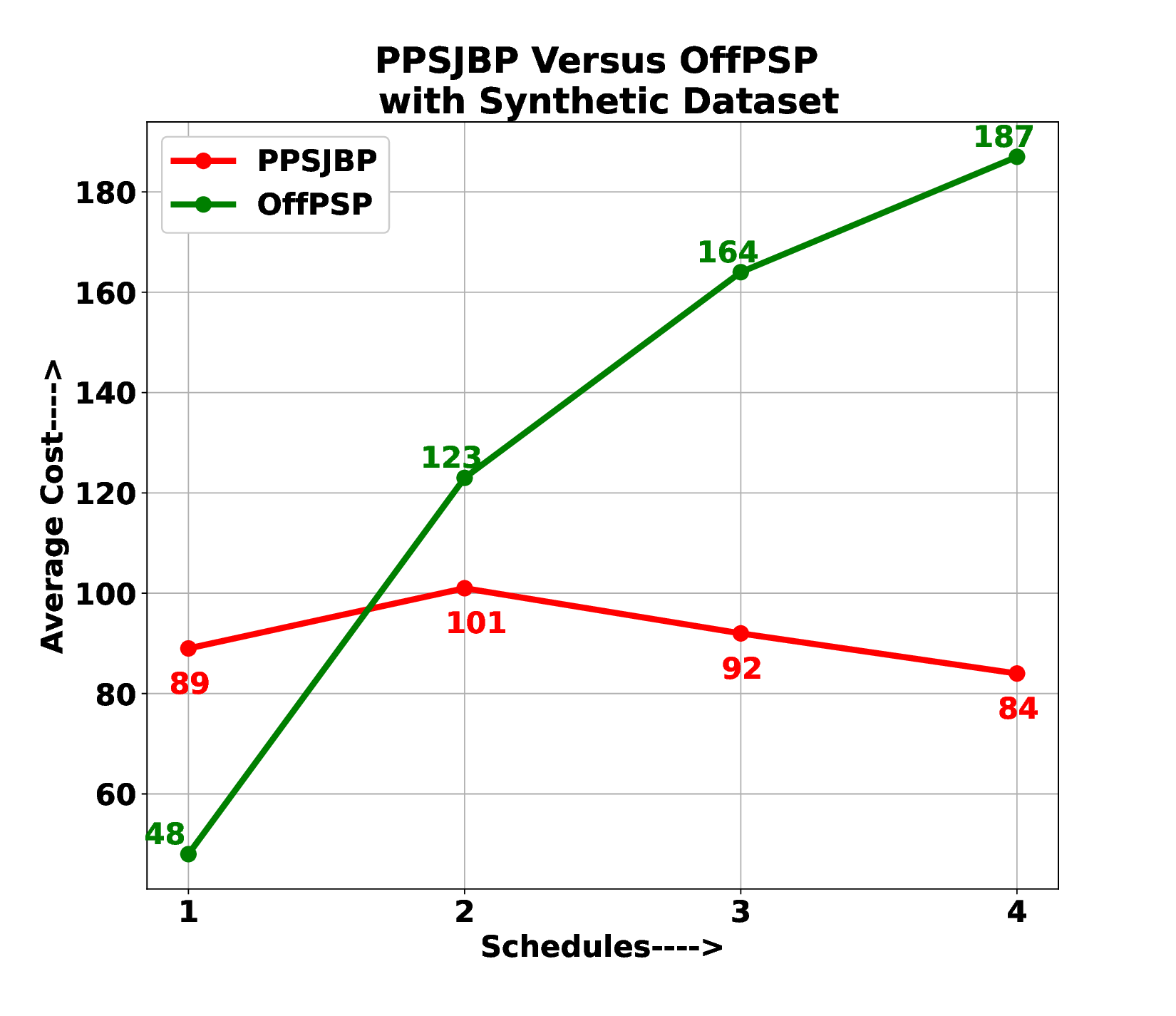}
    \label{bvo-1}}
  \end{subfloat}
  \begin{subfloat}[b][KDD Cup]
  {
    \includegraphics[width=0.30\linewidth]{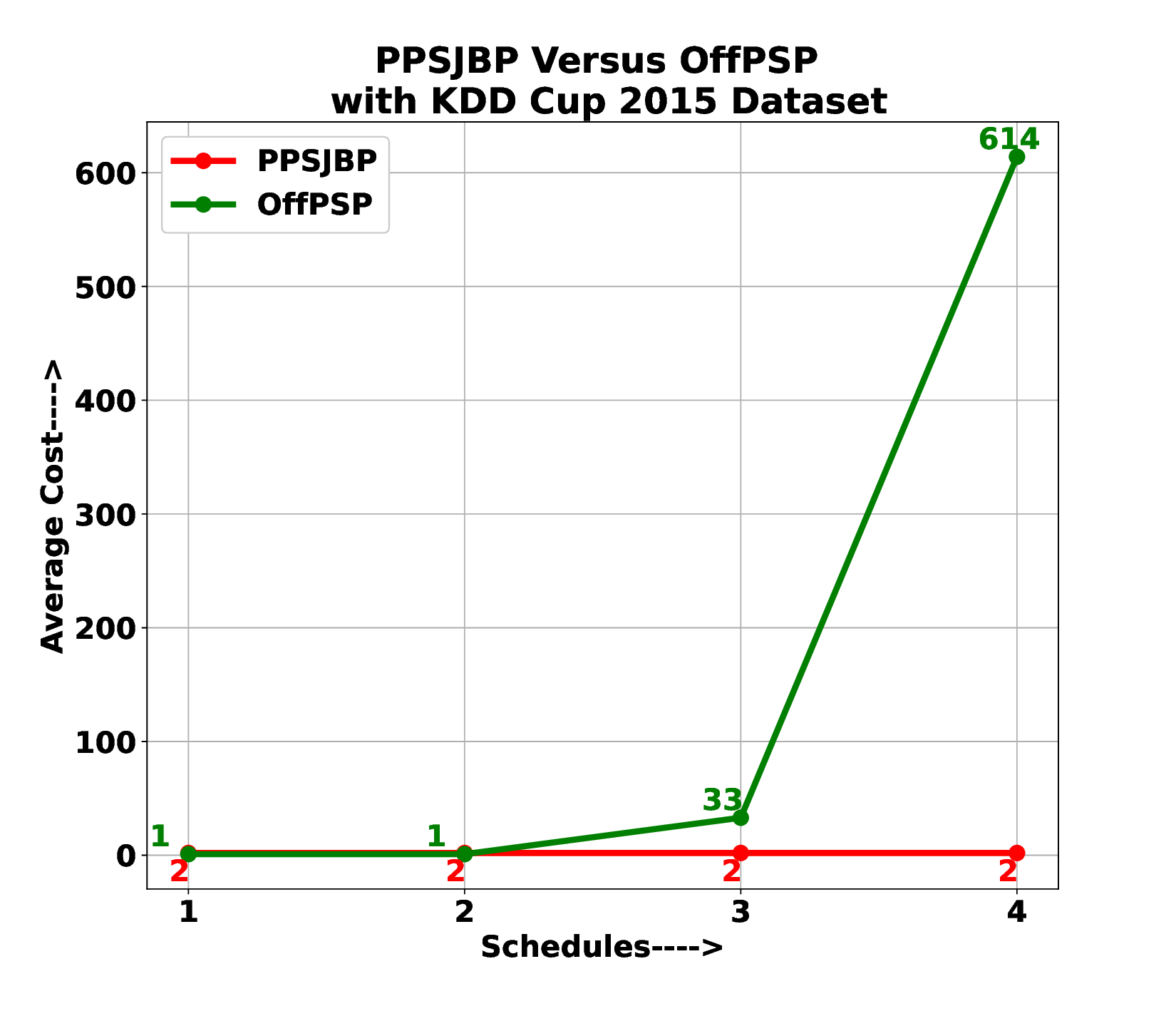}
    \label{bvo-2}}
  \end{subfloat}
  \begin{subfloat}[b][Bus Driver]
  {
    \includegraphics[width=0.30\linewidth]{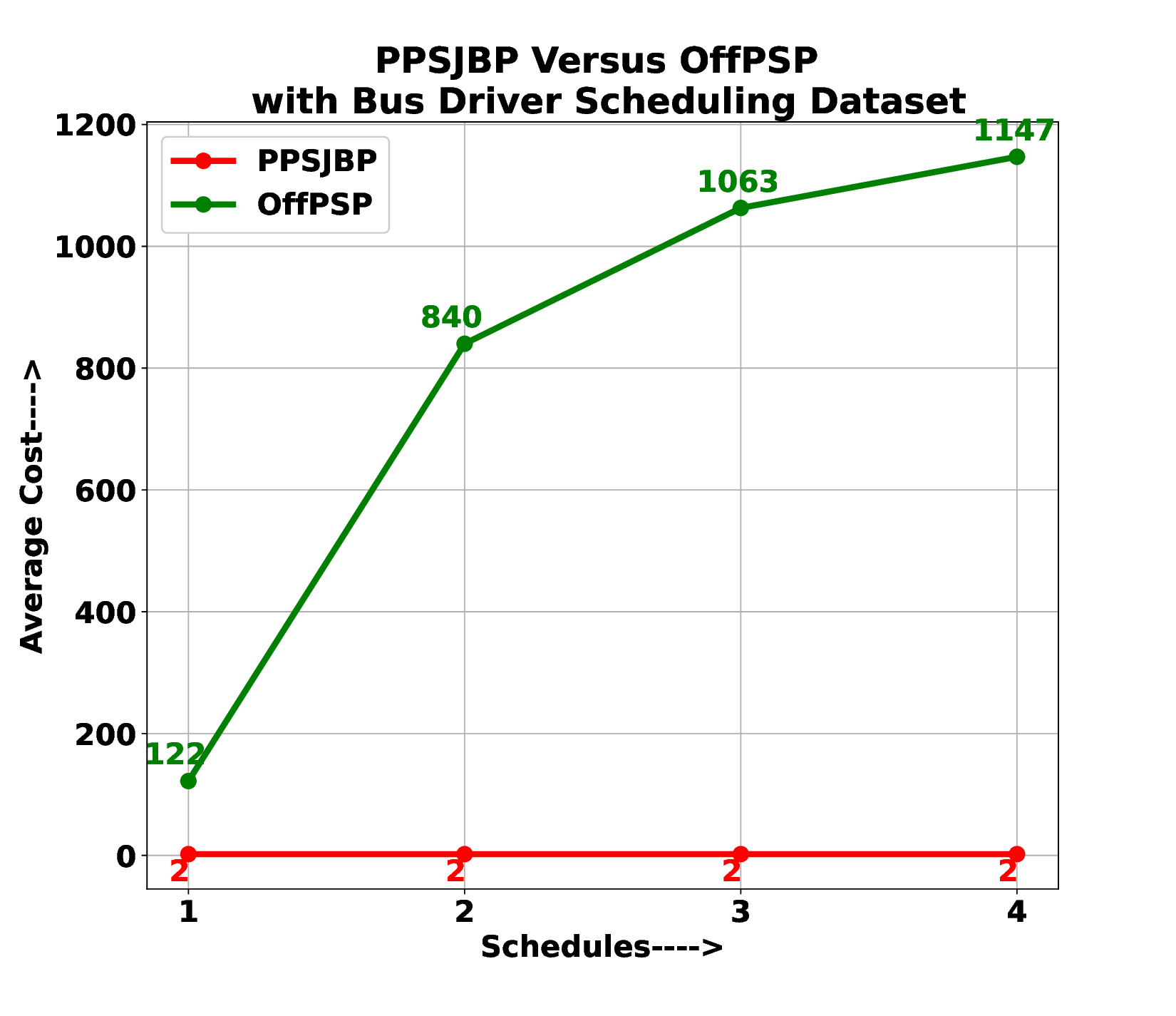}
    \label{bvo-3}}
  \end{subfloat}
  \caption{Comparison Between PPSJBP and OffPSP over Average Cost of each Schedule}
\end{figure}

In OffPSP algorithm, it was also mentioned that after allocation of jobs into a schedule, if the utility of the last job allocated into a schedule is greater than the utility of the other jobs, the last job is kept and the other jobs are discarded. \textit{Wang. et. al.} has not processed this discarded jobs further (as per the algorithm is presented in \cite{wang2019procrastination}). This creates a certain amount of disparity in the job allocation process. 
It is observed that the OffPSP may not perform efficiently in terms of balancing of the scheduled jobs as they are scheduling the jobs in a sorted order by using the concept of marginal utility and also, considering whether the utility of the of the last job allocated into a schedule is greater than the utility of
the other jobs.

\subsection{OffPSP Vs. PPSJBP}\label{ovp}
This comparative study will be done on the basis of various parameters which are: $i)$ Average cost of each schedule, $ii)$ Total number of jobs allocated in each schedule. The set up for simulation of OffPSP algorithm in the perspective of our work is given as: (1) Number of schedules: $|\Delta|=4$, (2) Deadline for each schedule=1 week, (3) Threshold for each schedule: $\delta_i^{Th}= \frac{\sum_{i=1}^{|\Delta|}\sum_{j=1}^{|\delta_i|}\chi_{ij}}{|\Delta|}$, and (4) Utility of $\gamma_i, \forall i=1$.

\subsubsection{OffPSP Vs. PPSJBP - On Average Cost per Schedule} 
With the above settings, the OffPSP and PPSJBP algorithm is simulated with the synthetic dataset with 200 jobs, Bus Driver Scheduling dataset with 359 jobs and KDD Cup 2015 dataset with more than 3000 jobs. The schedules produced by the OffPSP is unbalanced in terms of average cost of jobs than PPSJBP. A comparison graph is shown from \cref{bvo-1} to \cref{bvo-3} using the Synthetic dataset, KDD Cup 2015 dataset, and Bus Driver Scheduling dataset respectively. It is pretty evident from the figure that PPSJBP is behaving in most balanced way, whereas OffPSP is varying (increasing) over average cost of each schedule.
\begin{figure}
\centering
\begin{subfloat}[b][Synthetic]{
    \includegraphics[width=0.30\linewidth]{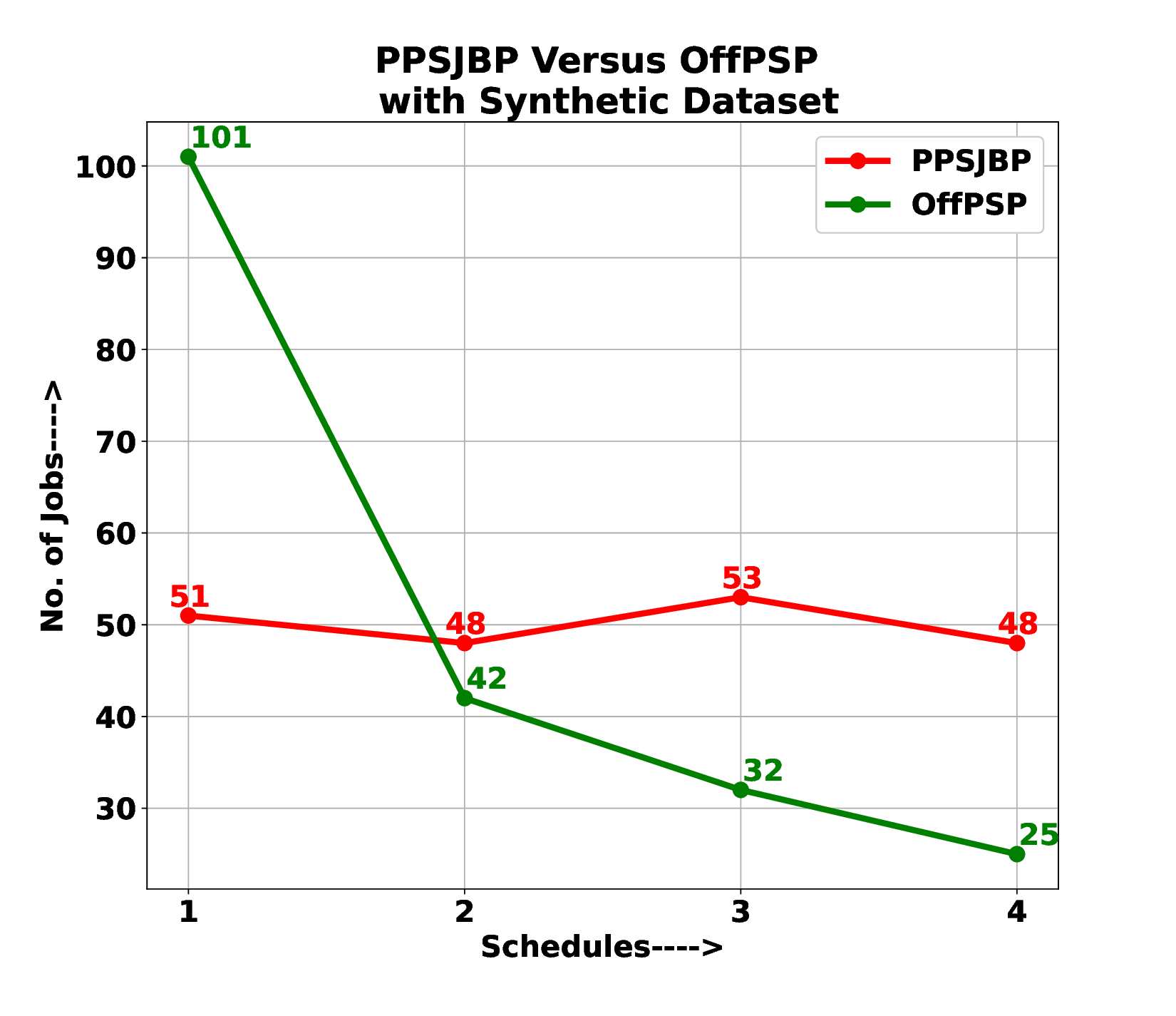}
    \label{bvo-4}}
  \end{subfloat}
  \begin{subfloat}[b][Bus Driver]
  {
    \includegraphics[width=0.30\linewidth]{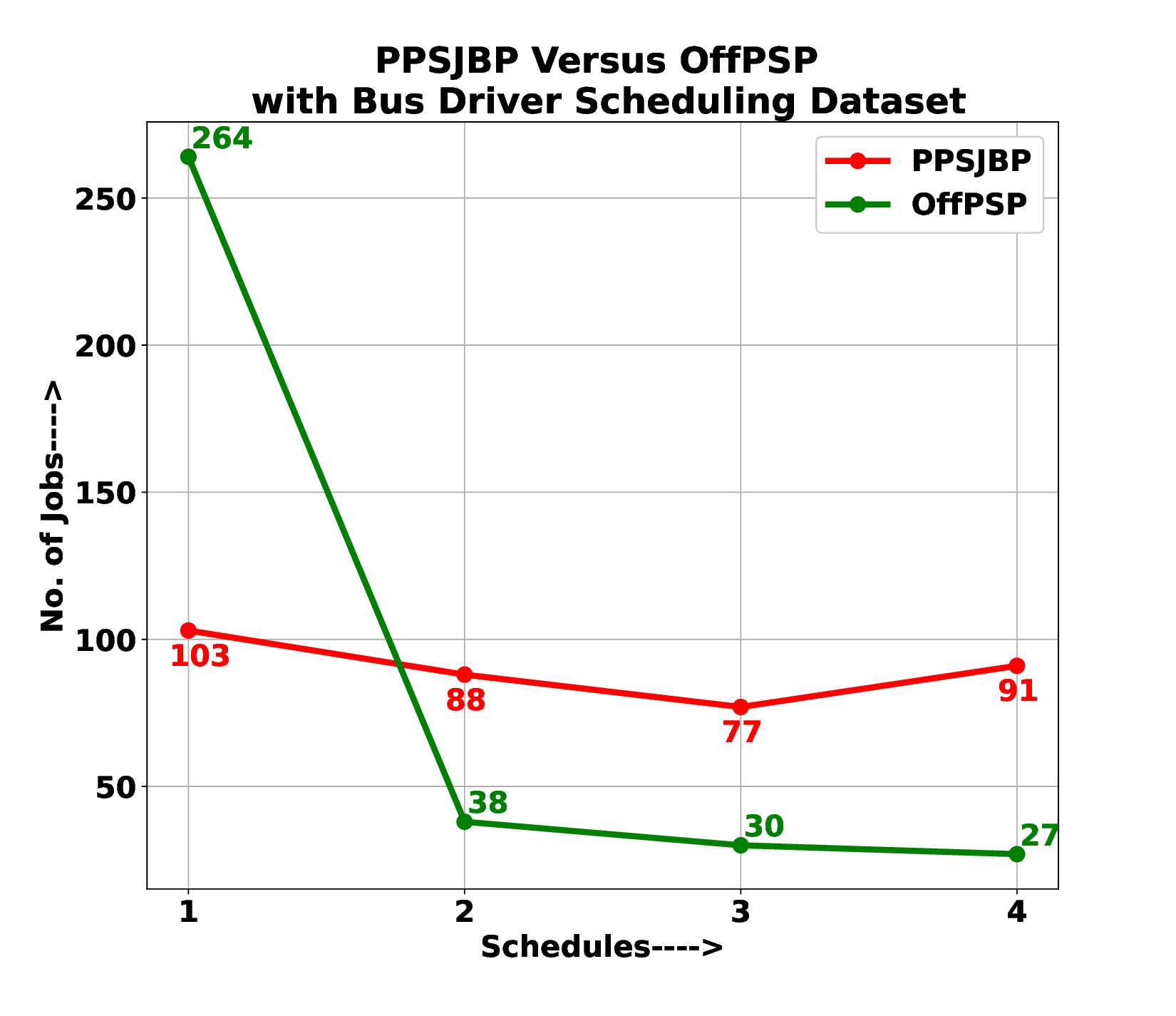}
    \label{bvo-5}}
  \end{subfloat}
  \begin{subfloat}[b][KDD Cup]
  {
    \includegraphics[width=0.30\linewidth]{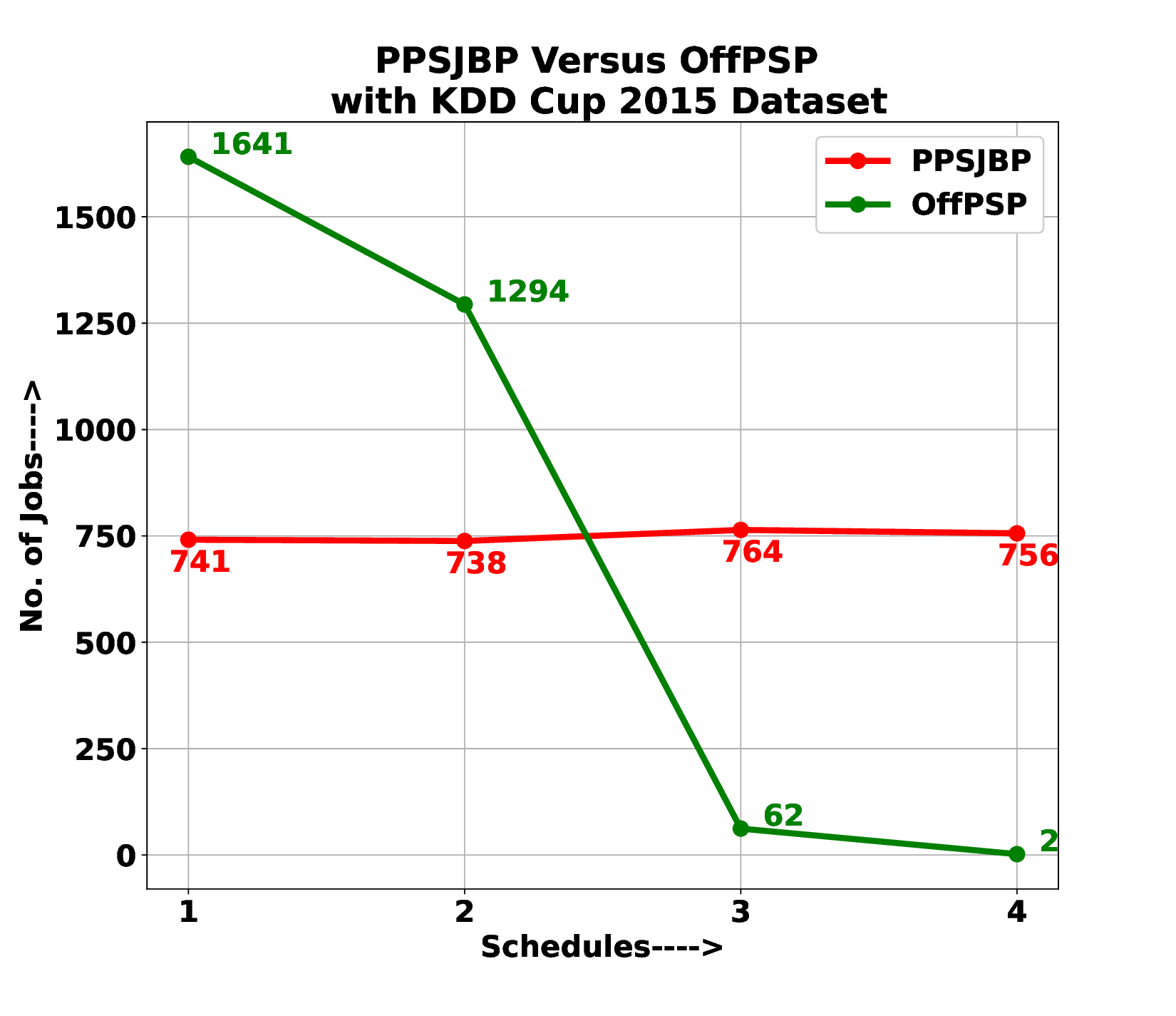}
    \label{bvo-6}}
  \end{subfloat}
  \caption{Comparison Between PPSJBP and OffPSP over Number of Jobs in Each Schedule}
\end{figure}

\subsubsection{OffPSP Vs. PPSJBP - On  Number of Jobs per Schedule}\label{ovbnj}
In this comparison, we will see that the number of jobs that are allocated into different schedules largely vary in OffPSP algorithm and are mostly balanced in PPSJBP algorithm. This is due to the fact that in OffPSP, all jobs are allocated into each schedule as per the decreasing order of the ratio of marginal utility and cost. In simple terms, all jobs are allocated into each schedule in sorted order. In our algorithm, all jobs are randomly allocated with large number of repetition. Accordingly, the schedules with minimum variance has produced the most balanced allocation. We have shown this comparison using Synthetic, Bus Driver Scheduling and KDD Cup 2015 datasets, from \cref{bvo-4} to \cref{bvo-6} respectively. As the number of jobs in each schedule is almost equal, the average cost of each schedule is almost uniform in our algorithm. It is due to the fact that higher cost jobs are distributed among all the schedules. According to \textit{OffPSP}, the number of jobs in each schedule largely vary. As a result the higher cost jobs are accumulated into one schedule which enhances the chance of procrastination.

In Lemma \ref{lem2}, we have shown that the average number of jobs to be assigned in each schedule is $\frac{|\Gamma|}{|\Delta|}$. In this section, in the above simulation we have found that all the schedules are allocated with almost equal number of jobs. This simulation, thus properly justifies Lemma \ref{lem2}.

\section{Conclusion and Future Works}\label{cnl}
In this paper we have proposed a novel procrastination preventive algorithm named PPSJBP which distributes jobs into different schedules (in most balanced manner) as per their costs to avoid procrastination in spatial crowdsourcing application. Every schedule is given equal time frame by subdividing the total time frame so that the on-time and efficient completion of jobs are guaranteed. Another aspect of PPSJBP is that, the schedules which are finally given to the agents are presented in non increasing order of their cost. The benefit of this is two fold: (1) the agents will have less burden as they will be going down the line to execute jobs in several schedules and (2) the task provider will have ample time to work on the little bit heavy schedule as it is submitted earlier. So the PPSJBP is relying on 3 facts: division of the time horizon into several schedules (choice reduction), aligning the jobs into the schedules by a randomized repetitive procedure (balancing), re-arrangement of the set of schedules (with minimum variance) in non increasing order of the cost of each schedule (benefiting task provider). We have
compared our scheme with the existing algorithm
with synthetic as well as real datasets
through extensive simulation and it is observed that in terms of balancing effect,
our proposed algorithm outperforms the existing one. Also we have shown analytically that our proposed algorithm maintains the
balanced distribution.

As it is a new paradigm for spatial crowdsourcing, there are multiple future directions to address the prevention of procrastination. In a SC zone, the distance between two distinct locations might be very less. In such situation, these two locations coulde be merged to address as a single location. Platform-centric data about the agents may be used to avoid procrastination in a better way. Another direction could be to give incentives to the agents on the way of performing the task to avoid procrastination.

\bibliography{Ref}

\end{document}